\documentclass[reprint,superscriptaddress,nofootinbib,amsmath,amssymb,floatfix,float,aps]{revtex4-2}
\usepackage{graphicx}% Include figure files
\usepackage{dcolumn}% Align table columns on decimal point
\usepackage{bm}% bold math
\usepackage{hyperref}
\usepackage{cleveref}
\begin{document}

\title[]{Gravitational waves from inspiraling black holes in quadratic
gravity}

\author{Matheus F. S. Alves}
 \email{matheus.s.alves@edu.ufes.br}
\affiliation{
Departamento de Física Teórica e Experimental,
Universidade Federal do Rio Grande do Norte, Natal, RN, 59078-970, Brazil.}
\affiliation{Departamento de Física, Universidade Federal do Espírito Santo, Vitória, ES,  29075-910, Brazil.}
\author{Luís F. M. A. M. Reis}%
 \email{luisfmamreis@gmail.com}
\affiliation{
Departamento de Física Teórica e Experimental,
Universidade Federal do Rio Grande do Norte, Natal, RN, 59078-970, Brazil.}%

\author{L. G. Medeiros}
 \email{leo.medeiros@ufrn.br}
\affiliation{%
Escola de Ci\^encia e Tecnologia, Universidade Federal do Rio Grande
do Norte, Campus Universit\'ario, s/n\textendash Lagoa Nova, CEP 59078-970,
Natal, Rio Grande do Norte, Brazil.}%

\begin{abstract}
 We perform a study of gravitational waves emitted by inspiraling black holes in the context of quadratic gravity. By linearizing the field equations around a flat background, we demonstrate that all degrees of freedom satisfy wavelike equations. These degrees of freedom split into three modes: a massive spin-$2$ mode, a massive spin-$0$ mode, and the expected massless spin-$2$ mode. We construct the energy-momentum tensor of gravitational waves and show that, due to the massive spin-$2$ mode, it presents the Ostrogradsky instability. We also show how to deal with this possible pathology and obtain consistent physical interpretations for the system. Using the energy-momentum tensor, we study the influence of each massive mode in the orbital dynamics and compare it with the standard result of general relativity. Moreover, we present two methods to constrain the parameter $\alpha$ associated with the massive spin-$2$ contribution. From the first method, using the combined waveform for the spin-$2$ modes, we obtain the constraint $ \alpha \lesssim 1.1 \times 10^{21} m^{2}$. In the second method, using the coalescence time, we get the constraint $ \alpha \lesssim 1.1 \times 10^{13} m^{2}$.
\end{abstract}

\maketitle

\affiliation{
Departamento de Física Teórica e Experimental,
Universidade Federal do Rio Grande do Norte, Natal, RN, 59078-970, Brazil.}
\affiliation{Departamento de Física, Universidade Federal do Espírito
Santo, Vitória, ES,  29075-910, Brazil.}

\affiliation{
Departamento de Física Teórica e Experimental,
Universidade Federal do Rio Grande do Norte, Natal, RN, 59078-970, Brazil.}

\affiliation{Escola de Ci\^encia e Tecnologia, Universidade Federal do Rio Grande
do Norte, Campus Universit\'ario, s/n\textendash Lagoa Nova, CEP 59078-970,
Natal, Rio Grande do Norte, Brazil.}

\section{Introduction}

The direct detections of gravitational waves (GWs) by the LIGO-VIRGO
collaboration, first in 2015 due to the collision of two black holes and
then in 2017 due to the collision of two neutron stars, is one of the main
results in the history of general relativity (GR) \cite%
{PhysRevLett.116.061102,PhysRevLett.119.161101,Abbott_2021}. These
detections not only observationally confirm one of the fascinating results
of GR but also make it possible to obtain a deeper understanding of the
gravitational interaction. In addition to the direct detections, GWs had
already been indirectly observed through neutron star binary systems. In the
1970s, Hulse and Taylor observed a decrease in the orbital period of the PSR
1913+16 system \cite{hulse1975discovery}, entirely consistent with the
energy and momentum losses predicted by GR theory due to GWs emission \cite%
{PhysRev.166.1263,PhysRev.166.1272}.

Despite these and several other experimental results supporting the GR,
there is a consensus in the scientific community that it is an incomplete
theory. In addition to some structural problems within the theory, such as
the presence of singularities, the GR also presents problems when analyzed
in the context of high-energy physics. general relativity cannot be
quantized through the conventional methods of particle physics \cite%
{doi:10.1063/1.1724264,Woodard_2009}, and it struggles to describe the early
(inflationary) universe consistently \cite{STAROBINSKY198099,PhysRevD.23.347}%
.

One of the main approaches to solving these problems is modifying the GR,
especially the Einstein-Hilbert action. The first proposal to modify GR, and
in turn the simplest, came from Einstein himself, who introduced the
cosmological constant $\Lambda$ in theory as a way of solving the
cosmological view of his time \cite{gamow1956evolutionary}. Despite being
one of the main candidates for dark energy, this modification still faces
some observational problems that may even be associated with the current "cosmological crisis" \cite{Di_Valentino_2021,PhysRevD.103.L041301,Di_Valentino_2019,PhysRevD.96.043503,PhysRevD.102.023518,PhysRevD.101.063502,Vagnozzi_2021,Vagnozzi_20212}.

There are, however, more sophisticated modification proposals which involve
the addition of the curvature scalar terms, such as the $f(R)$ theories \cite%
{Sotiriou_2010,De_Felice_2010,Nojiri_2011,Capozziello_2011}. In this
scenario, the theory $f(R)=R+\alpha R%
$, for example, constitutes the so-called Starobinsky inflationary model
\cite{Gurovich:1979xg,STAROBINSKY198099}, which proves to be a consistent
approach to inflation \cite{2020}. Besides $f(R)$ theories, there are
proposals for several other modifications to the Einstein-Hilbert action,
e.g. theories that include a vector field \cite{PhysRevD.64.024028,PhysRevD.70.083509,Moffat_2006} and theories involving the addition of higher-order terms like $\nabla R$, $%
\nabla^{n}R$ and $R\square R$ \cite{PhysRevD.93.124034,PhysRevD.99.084053,asorey1997some}.
Therefore, the search for observational constraints in modified theories of
gravity proves to be fundamental in attempting to narrow down the parameters
of these theories.

One of the main ways of establishing observational constraints in theories
of modified gravity is in the cosmological context \cite%
{NOJIRI20171,Cuzinatto_2019,PhysRevD.105.063504,MS}. However, with the new
era of GWs detection, the analysis of modified theories in contexts of
gravitational waves \cite{Neto_2003,capozziello2019gravitational,Inagaki_2020,Yang_2011,H_lscher_2019,Caprini_2018} is also of great interest, even allowing the
determination of parameters in new scales different from those probed in
cosmology \cite{PhysRevD.84.024027,PhysRevD.85.044022,Capozziello_2010,PhysRevD.104.084061,pqg}.

In this spirit, we analyze the quadratic gravity model given by the action (%
\ref{acao0}) in the context of the emission of gravitational waves:
\footnote{%
These action was proposed by Stelle \cite%
{PhysRevD.16.953,Stelle:1977ry} to deal with the renormalizability problem
of the gravitational field.}

\begin{eqnarray}
S &=&\int d^{4}x\sqrt{-g}\left[ R+\frac{1}{2}\gamma R^{2}-\frac{1}{2}\alpha
C_{\mu \nu \alpha \beta }C^{\mu \nu \alpha \beta }\right]  \label{acao0} \\
&&-2\kappa \int d^{4}x\sqrt{-g}\mathcal{L}_{m},  \notag
\end{eqnarray}%
where $\kappa =\frac{8\pi G}{c^{4}}$, $\alpha $, $\gamma $ are coupling
constants with the dimension of (length)$^{2}$ and $C_{\mu \nu \alpha \beta
} $ is the Weyl tensor.

Quadratic gravity is embedded in a context of higher-order gravity
theories where higher energy corrections are inserted in the action. In this
context, the Einstein-Hilbert term and the cosmological constant are
zero-order terms (squared mass terms), the invariants $R^{2}$ and $C_{\mu
\nu\alpha\beta}C^{\mu\nu\alpha\beta}$ are first-order corrections\footnote{The
other two first-order terms, namely $\square R$ and $G=R^{2}-4R_{\mu\nu}R^{\mu\nu}+R_{\mu\nu\alpha\beta
}R^{\mu\nu\alpha\beta}$, do not contribute to the field equations.} (fourth
mass terms), quantities of the type $R\square R$, $C_{\mu\nu\alpha\beta
}\square C^{\mu\nu\alpha\beta}$ plus cubic terms in the Riemann tensor are
second-order corrections (sixth mass terms), etc.\footnote{More details about
this construction can be found in the introduction of \cite{arxiv.2207.02103}.} Thus, this
work aims to study how first-order corrections to general relativity influence
the emission of GWs emitted by binary systems. More specifically,
we study the emission of GWs produced by the inspiral phase of a binary black hole
system in the approximations of point masses, circular orbit, and nonrelativistic dynamics.

The paper has the following structure: in Sec. \ref{sec - weak field}%
, we carry out the linearization of the field equation decomposing the
metric into three new fields: a massless spin-$2$ field, as in GR; a massive
spin-$2$ field, and a massive spin-$0$ field. Next, we analyze the gauge
conditions that allow us to reduce the degrees of freedom of these new
fields significantly. In the Sec. \ref{sec solutions}, we obtain the
Green's functions for the massive fields and carefully analyze the
multipolar expansion for each field. In Sec. \ref{sec binary}, the
solutions of the fields in terms of the multipolar expansions are applied to
a binary system of point masses in circular orbits and the power radiated by
the GWs is obtained. With that, in Sec. \ref{sec inspiral}, the
effect of the emission of GWs is analyzed using the energy balance equation,
so that the inspiral phase is studied in terms of the orbital frequency of
the system. In addition, Ostrogradsky's instability, generated by the
Weyl-Weyl term, is carefully analyzed in this section, and then it is shown
how this apparent pathology can be suppressed and corrected. In light of
this apparent pathology, we discuss in Sec. \ref{sec spin 2} the role
that spin-$2$ fields play in the emission of GWs, mainly explaining the
differences and similarities between these fields, such as their propagation
velocities and waveforms. With this, we establish some observational constraints.
Finally, we end with the final comments in the Sec. \ref{sec final}.

\begin{widetext}
\section{Quadratic gravity in the weak-field regime\label{sec - weak field}}
The field equation is obtained from (\ref{acao0}) by varying it with respect
to $g^{\mu\nu}$:%
\begin{equation}
R_{\mu\nu}-\frac{1}{2}g_{\mu\nu}R+\gamma\left[  R\left(  R_{\mu\nu}-\frac
{1}{4}Rg_{\mu\nu}\right)  +g_{\mu\nu}\nabla_{\rho}\nabla^{\rho}R-\nabla_{\mu
}\nabla_{\nu}R\right]  -\alpha\left[  \nabla^{\rho}\nabla^{\beta}C_{\mu\rho
\nu\beta}+\frac{1}{2}R^{\rho\beta}C_{\mu\rho\nu\beta}\right]  =\kappa
T_{\mu\nu} \label{Eq Field}%
\end{equation}
with the Riemann and Weyl tensors defined as%
\begin{align}
R_{\text{ \ }\nu\alpha\beta}^{\kappa}  &  \equiv\partial_{\alpha}%
\Gamma_{\text{ \ }\nu\beta}^{\kappa}-\partial_{\beta}\Gamma_{\text{ \ }%
\nu\alpha}^{\kappa}+\Gamma_{\text{ \ }\rho\alpha}^{\kappa}\Gamma_{\text{
\ }\nu\beta}^{\rho}-\Gamma_{\text{ \ }\rho\beta}^{\kappa}\Gamma_{\text{ \ }%
\nu\alpha}^{\rho},\label{Tensor de Riemann}\\
C_{\mu\nu\alpha\beta}  &  \equiv R_{\mu\nu\alpha\beta}-\frac{1}{2}\left(
g_{\mu\alpha}R_{\beta\nu}-g_{\mu\beta}R_{\alpha\nu}+g_{\nu\beta}R_{\alpha\mu
}-g_{\nu\alpha}R_{\beta\mu}\right)  +\frac{1}{6}R\left(  g_{\mu\alpha}%
g_{\beta\nu}-g_{\mu\beta}g_{\alpha\nu}\right)  . \label{Tensor de Weyl}%
\end{align}
\end{widetext}
In addition, the trace of the field equation (\ref{Eq Field}) is%
\begin{equation}
3\gamma\nabla_{\rho}\nabla^{\rho}R-R=\kappa T.  \label{Eq Field Traco}
\end{equation}

The linearized equations on a flat background for quadratic gravity were
obtained in Ref. \cite{Teyssandier_1989}. However, we will retrieve these
equations using a clearer and more simplified approach. We start by
decomposing the metric as\footnote{%
We adopted the $\left( -1,1,1,1\right) $ signature for the metric.}%
\begin{equation*}
g_{\mu \nu }=\eta _{\mu \nu }+h_{\mu \nu },
\end{equation*}%
where $\left\vert h_{\mu \nu }\right\vert <<1$. From the definition of the
trace-reverse tensor $h_{\nu \beta }=\bar{h}_{\nu \beta }-\frac{1}{2}\eta
_{\nu \beta }\bar{h}$, we obtain in linear order
\begin{align}
C_{\mu \nu \alpha \beta }^{\left( 1\right) }& =\frac{1}{2}\left[ \partial
_{\alpha }\partial _{\nu }\bar{h}_{\mu \beta }-\partial _{\beta }\partial
_{\nu }\bar{h}_{\mu \alpha }+\partial _{\beta }\partial _{\mu }\bar{h}_{\nu
\alpha }-\partial _{\alpha }\partial _{\mu }\bar{h}_{\nu \beta }\right]
\notag \\
& -\frac{1}{4}\eta _{\mu \alpha }\left( \partial _{\rho }\partial _{\nu }%
\bar{h}_{\text{ \ }\beta }^{\rho }-\partial _{\beta }\partial _{\nu }\bar{h}%
+\partial _{\beta }\partial _{\rho }\bar{h}_{\nu }^{\text{ \ }\rho }-\square
\bar{h}_{\nu \beta }\right)  \notag \\
& +\frac{1}{4}\eta _{\mu \beta }\left( \partial _{\rho }\partial _{\nu }\bar{%
h}_{\text{ \ }\alpha }^{\rho }-\partial _{\alpha }\partial _{\nu }\bar{h}%
+\partial _{\alpha }\partial _{\rho }\bar{h}_{\nu }^{\text{ \ }\rho
}-\square \bar{h}_{\nu \alpha }\right)  \notag \\
& -\frac{1}{4}\eta _{\nu \beta }\left( \partial _{\rho }\partial _{\mu }\bar{%
h}_{\text{ \ }\alpha }^{\rho }-\partial _{\alpha }\partial _{\mu }\bar{h}%
+\partial _{\alpha }\partial _{\rho }\bar{h}_{\mu }^{\text{ \ }\rho
}-\square \bar{h}_{\mu \alpha }\right)  \notag \\
& +\frac{1}{4}\eta _{\nu \alpha }\left( \partial _{\rho }\partial _{\mu }%
\bar{h}_{\text{ \ }\beta }^{\rho }-\partial _{\beta }\partial _{\mu }\bar{h}%
+\partial _{\beta }\partial _{\rho }\bar{h}_{\mu }^{\text{ \ }\rho }-\square
\bar{h}_{\mu \beta }\right)  \notag \\
& +\frac{1}{6}\left[ \partial _{\sigma }\partial _{\rho }\bar{h}^{\sigma
\rho }-\square \bar{h}\right] \left( \eta _{\mu \alpha }\eta _{\beta \nu
}-\eta _{\mu \beta }\eta _{\alpha \nu }\right) ,  \label{Weyl transverso}
\end{align}%
and
\begin{align}
R_{\mu \nu }^{\left( 1\right) }& \approx \frac{1}{2}\left[ \partial _{\rho
}\partial _{\nu }\bar{h}_{\mu }^{\text{ \ }\rho }+\partial _{\rho }\partial
_{\mu }\bar{h}_{\nu }^{\text{ \ }\rho }-\square \bar{h}_{\mu \nu }+\frac{1}{2%
}\eta _{\mu \nu }\square \bar{h}\right] ,  \label{Ricci transverso} \\
R^{\left( 1\right) }& \approx \partial _{\sigma }\partial _{\rho }\bar{h}%
^{\sigma \rho }+\frac{1}{2}\square \bar{h}
\label{Escalar de curvatura transverso}
\end{align}%
where $\square \equiv \partial _{\alpha }\partial ^{\alpha }$ and $\bar{h}%
\equiv \bar{h}_{\text{ \ }\alpha }^{\alpha }$.

The objective is to obtain wave equations for the various degrees of freedom
of the metric in quadratic gravity. We start by defining the dimensionless
scalar quantity $\Phi \equiv -\gamma R^{\left( 1\right) }$. In this case,
the linearized equation (\ref{Eq Field Traco}) can be written as%
\begin{equation}
\left( \square -m_{\Phi }^{2}\right) \Phi =-\frac{\kappa T}{3},
\label{Eq field lin 3}
\end{equation}%
where $m_{\Phi }^{2}=1/\left( 3\gamma \right) $ and $T$ is the trace of $%
T_{\mu \nu }$ in its linearized form. Then we decompose the metric into a
scalar part and a tensorial part \cite{teyssandier1990new,H_lscher_20192}
\begin{equation}
\bar{h}_{\mu \nu }=\Theta _{\mu \nu }-\eta _{\mu \nu }\left( \Phi +\phi
\right)  \label{decomposicao 1}
\end{equation}%
with $\phi $ representing an auxiliary gauge field\footnote{%
The $\Phi $ field contains the entire scalar degree of freedom of the metric.%
}. Furthermore, for the spin-$2$ part, we adopt the gauge%
\begin{equation}
\partial ^{\nu }\Theta _{\mu \nu }=0\Rightarrow \partial ^{\nu }\bar{h}_{\mu
\nu }=-\partial _{\mu }\left( \Phi +\phi \right) .  \label{Gauge}
\end{equation}%
Substituting the decomposition (\ref{decomposicao 1}) in the linearized
equation (\ref{Eq Field}) we get%
\begin{align}
\square\Theta_{\mu\alpha} & +2\partial_{\mu}\partial_{\alpha}\phi-2\eta
_{\mu\alpha}\square\phi\label{Eq Field linearized Aux}\\ \nonumber
& +2\alpha\partial^{\nu}\partial^{\beta}C_{\mu\nu\alpha\beta}^{\left(  1\right)
}=-2\kappa T_{\mu\alpha}.
\end{align}

The next step is to separate the tensorial component into a massive and a
nonmassive mode:%
\begin{equation}
\Theta_{\mu\alpha}=\tilde{h}_{\mu\alpha}+\Psi_{\mu\alpha},
\label{decomposicao 2}
\end{equation}
where we impose that the nonmassive mode $\tilde{h}_{\mu\alpha}$ satisfies
the equation
\begin{equation}
\square\tilde{h}_{\mu\alpha}=-2\kappa T_{\mu\alpha}.  \label{Eq field lin 1}
\end{equation}
Thus, Eq. (\ref{Eq Field linearized Aux}) can be rewritten as%
\begin{equation}
\square\Psi_{\mu\alpha}+2\partial_{\mu}\partial_{\alpha}\phi-2\eta_{\mu%
\alpha
}\square\phi+2\alpha\partial^{\nu}\partial^{\beta}C_{\mu\nu\alpha\beta
}^{\left( 1\right) }=0.  \label{Eq Aux spin 2 massivo}
\end{equation}
\begin{widetext}
Then, we use the decomposition of the metric in the Weyl tensor and calculate
$\partial^{\nu}\partial^{\beta}C_{\mu\nu\alpha\beta}^{\left(  1\right)  }$
obtaining%
\[
\partial^{\nu}\partial^{\beta}C_{\mu\nu\alpha\beta}^{\left(  1\right)
}=-\frac{1}{4}\square^{2}\left(  \tilde{h}_{\mu\alpha}+\Psi_{\mu\alpha
}\right)  +\frac{1}{12}\eta_{\mu\alpha}\square^{2}\left(  \tilde{h}%
+\Psi\right)  -\frac{1}{12}\partial_{\alpha}\partial_{\mu}\square\left(
\tilde{h}+\Psi\right)  ,
\]
where $\tilde{h}=\tilde{h}_{\alpha\text{ }}^{\alpha}$ and $\Psi=\Psi_{\alpha
}^{\alpha}$.
\end{widetext}
Once $\partial^{\nu}\partial^{\beta}C_{\mu\nu\alpha\beta}^{\left( 1\right) }$
has been calculated, we substitute this result in (\ref{Eq Aux spin 2
massivo}) and choose the gauge field $\phi$ as\footnote{%
This choice is necessary to obtain a wave equation for $\Psi _{\mu\alpha}$.}
\begin{equation}
\phi=\frac{\alpha}{3}\square\left( \tilde{h}+\Psi\right) .  \label{Def phi}
\end{equation}
So, Eq. (\ref{Eq Aux spin 2 massivo}) is rewritten as%
\begin{equation}
\square\left[ \Psi_{\mu\alpha}-\frac{\alpha}{2}\square\left( \tilde{h}%
_{\mu\alpha}+\Psi_{\mu\alpha}\right) \right] =0.
\label{Eq spin 2 massive aux}
\end{equation}
For trivial boundary conditions (e.g. fields vanishing at infinity), the
above equation is satisfied only if the term in square brackets is null. In
this case, using Eq. (\ref{Eq field lin 1}) we get%
\begin{equation}
\left( \square-m_{\Psi}^{2}\right) \Psi_{\mu\alpha}=2\kappa T_{\mu\alpha},
\label{Eq field lin 2}
\end{equation}
where $m_{\Psi}^{2}=2/\alpha.$ From the field equations (\ref{Eq field lin 1}%
) and (\ref{Eq field lin 2}) we can show that the gauge conditions (\ref%
{Gauge}) and (\ref{Def phi}) result in%
\begin{equation}
\partial^{\mu}\Psi_{\mu\nu}=0\text{, \ }\partial^{\mu}\tilde{h}_{\mu\nu }=0%
\text{ \ and \ }\phi=\frac{1}{3}\Psi.  \label{Condicao de gauge final}
\end{equation}

Therefore, the decomposition of the metric in the form%
\begin{equation}
\bar{h}_{\mu\nu}=\tilde{h}_{\mu\nu}+\Psi_{\mu\nu}-\eta_{\mu\nu}\left( \Phi+%
\frac{1}{3}\Psi\right) ,  \label{Decomposicao metrica}
\end{equation}
and the choice of gauge (\ref{Condicao de gauge final}) result in the
Eqs. (\ref{Eq field lin 3}), (\ref{Eq field lin 1}) and (\ref{Eq field
lin 2}) associated with massive spin-$0$, massless spin-$2$ and massive spin-%
$2$ degrees of freedom, respectively.

\subsection{Degrees of freedom\label{sec - DoF}}

As in general relativity, the conditions given in (\ref{Condicao de gauge
final}) do not completely fix the gauge. Performing an infinitesimal
coordinate transformation%
\begin{equation*}
x^{\prime\mu}=x^{\mu}+\xi^{\mu}\left( x\right) ,
\end{equation*}
it is possible to show that the full spin-$2$ mode transforms as%
\begin{equation}
\Theta_{\mu\nu}^{\prime}=\Theta_{\mu\nu}-\partial_{\mu}\xi_{\nu}-\partial
_{\nu}\xi_{\mu}+\eta_{\mu\nu}\partial^{\beta}\xi_{\beta}.
\label{trans tensor Tr reverso}
\end{equation}
Taking the divergence of the expression above, we notice that any
transformation $\xi_{\nu}$ that respects $\square\xi_{\nu}=0$ maintains the
harmonic gauge $\partial^{\mu}\Theta_{\mu\nu}=0$. This residual gauge
freedom allows us to choose a transformation in the form%
\begin{equation}
\xi_{\nu}=b_{\nu}e^{i\bar{k}_{\alpha}x^{\alpha}}+b_{\nu}^{\ast}e^{-i\bar {k}%
_{\alpha}x^{\alpha}},  \label{xi}
\end{equation}
where $\bar{k}_{\alpha}\bar{k}^{\alpha}=0$ and $b_{\nu}$ is an arbitrary
constant vector.

The next step is to show that this extra degree of freedom, incorporated in
the $b_{\nu }$ vector, acts only in massless spin-$2$ modes. This can be
verified from the vacuum solutions of Eqs. (\ref{Eq field lin 1})
and (\ref{Eq field lin 2})%
\begin{equation*}
\tilde{h}_{\mu \nu }=\tilde{\varepsilon}_{\mu \nu }e^{ik_{\alpha }x^{\alpha
}}+\tilde{\varepsilon}_{\mu \nu }^{\ast }e^{-ik_{\alpha }x^{\alpha }}
\end{equation*}%
and%
\begin{equation*}
\Psi _{\mu \nu }=\varepsilon _{\mu \nu }e^{iq_{\alpha }x^{\alpha
}}+\varepsilon _{\mu \nu }^{\ast }e^{-iq_{\alpha }x^{\alpha }}\text{,}
\end{equation*}%
where $k_{\alpha }k^{\alpha }=0$ and $q_{\alpha }q^{\alpha }=-m_{\Psi }^{2}$%
. Substituting this last result and Eq. (\ref{xi}) in the transformation (%
\ref{trans tensor Tr reverso}) we get%
\begin{align}
\varepsilon _{\mu \nu }^{\prime }& =\varepsilon _{\mu \nu },  \notag \\
\tilde{\varepsilon}_{\mu \nu }^{\prime }& =\tilde{\varepsilon}_{\mu \nu
}+k_{\mu }b_{\nu }+k_{\nu }b_{\mu }-\eta _{\mu \nu }k^{\beta }b_{\beta }.
\label{gauge residual}
\end{align}

Based on Eqs. (\ref{Condicao de gauge final}) and (\ref{gauge residual}), we
conclude that the massless spin-$2$ mode has only $2$ degrees of freedom as
in general relativity. Furthermore, for the vacuum solution, we can choose
the traceless-transverse gauge where $\tilde{h}_{\mu0}=0$, $\partial^{i}%
\tilde {h}_{ij}=0$ and $\tilde{h}_{i}^{i}=0$. The massive spin-$2$ mode has
$5$ degrees of freedom since $\partial^{\mu}\Psi_{\mu\nu}=0$ and $%
\Psi=3\phi $. Considering the vacuum solution we can choose $\phi=\Psi/3=0$.

For completeness, it is interesting to explicitly determine the $5$ degrees
of freedom of $\Psi _{\mu \nu }$ for a monochromatic wave propagating in the
$\mathrm{z}$-direction \cite{BOGDANOS2010236}. In this case,%
\begin{equation*}
q^{\mu }=\left( \omega _{q},0,0,q\right) \text{ \ and \ }\omega
_{q}^{2}=q^{2}+m_{\Psi }^{2}.
\end{equation*}%
Using the gauge conditions $\partial ^{\mu }\Psi _{\mu \nu }=0$, $\Psi =0$
and defining%
\begin{align}
\Psi _{+}& =\frac{\Psi _{11}-\Psi _{22}}{2}\text{, \ }\Psi _{D}=\Psi
_{11}+\Psi _{22}\text{,}  \notag \\
\Psi _{\times }& =\Psi _{12}\text{, \ }\Psi _{B}=-\Psi _{13}\text{ \ and \ }%
\Psi _{C}=-\Psi _{23}  \label{Def Psi matrix}
\end{align}%
\begin{widetext}
we get
\begin{align*}
\Psi_{00}  &  =-\frac{q^{2}}{m_{\Psi}^{2}}\Psi_{D}\text{, }\Psi_{01}=\frac
{q}{\omega_{q}}\Psi_{B}\text{, }\Psi_{02}=\frac{q}{\omega_{q}}\Psi_{C}\text{,}
\Psi_{03} =\frac{q\omega_{q}}{m_{\Psi}^{2}}\Psi_{D}\text{, }\Psi_{33}%
=-\frac{\omega_{q}^{2}}{m_{\Psi}^{2}}\Psi_{D}, \\
\Psi_{11}  &  =\frac{1}{2}\Psi_{D}+\Psi_{+}\text{ \ and \ }\Psi_{22}=\frac
{1}{2}\Psi_{D}-\Psi_{+}\text{.}%
\end{align*}
Therefore, in matrix form, we have
\begin{align*}
\Psi_{\mu\nu}  & =\Psi_{+}\left(
\begin{array}
[c]{cccc}%
0 & 0 & 0 & 0\\
0 & 1 & 0 & 0\\
0 & 0 & -1 & 0\\
0 & 0 & 0 & 0
\end{array}
\right)  +\Psi_{\times}\left(
\begin{array}
[c]{cccc}%
0 & 0 & 0 & 0\\
0 & 0 & 1 & 0\\
0 & 1 & 0 & 0\\
0 & 0 & 0 & 0
\end{array}
\right)  +\Psi_{B}\left(
\begin{array}
[c]{cccc}%
0 & \frac{q}{\omega_{q}} & 0 & 0\\
\frac{q}{\omega_{q}} & 0 & 0 & -1\\
0 & 0 & 0 & 0\\
0 & -1 & 0 & 0
\end{array}
\right)  +\Psi_{C}\left(
\begin{array}
[c]{cccc}%
0 & 0 & \frac{q}{\omega_{q}} & 0\\
0 & 0 & 0 & 0\\
\frac{q}{\omega_{q}} & 0 & 0 & -1\\
0 & 0 & -1 & 0
\end{array}
\right)  \\
& +\Psi_{D}\left(
\begin{array}
[c]{cccc}%
-\frac{q^{2}}{m_{\Psi}^{2}} & 0 & 0 & \frac{q\omega_{q}}{m_{\Psi}^{2}}\\
0 & \frac{1}{2} & 0 & 0\\
0 & 0 & \frac{1}{2} & 0\\
\frac{q\omega_{q}}{m_{\Psi}^{2}} & 0 & 0 & -\frac{\omega_{q}^{2}}{m_{\Psi}%
^{2}}%
\end{array}
\right)  ,
\end{align*}
$\allowbreak$where $\Psi_{+}$, $\Psi_{\times}$, $\Psi_{B}$, $\Psi_{C}$ and
$\Psi_{D}$ correspond to the $5$ degrees of freedom of $\Psi_{\mu\nu}$.
\end{widetext}

\section{Solutions to the linearized field equations\label{sec solutions}}

The physical solution of Eq. (\ref{Eq field lin 1}) is well known and given
by%
\begin{equation*}
\tilde{h}_{\mu\nu}\left( t_{r},\mathbf{x}\right) =\frac{\kappa}{2\pi}\int
d^{3}x^{\prime}\frac{1}{\left\vert \mathbf{x}-\mathbf{x}^{\prime}\right\vert
}T_{\mu\nu}\left( t_{r},\mathbf{x}^{\prime}\right) ,
\end{equation*}
where $t_{r}=t-\frac{\left\vert \mathbf{x}-\mathbf{x}^{\prime}\right\vert }{c%
}$ is the retarded time and $\mathbf{x}$ (the vector $\mathbf{x}^{\prime}$)
points from the origin of the coordinate system to the observer (to the
source). The formal solutions to the equations (\ref{Eq field lin 2}) and (%
\ref{Eq field lin 3}) are given by%
\begin{align}
\Phi\left( \mathbf{x},t\right) & =-\frac{\kappa}{3}\int G_{\Phi}\left(
x^{\mu};x^{\mu^{\prime}}\right) T\left( t^{\prime},\mathbf{x}^{\prime
}\right) d^{4}x^{\prime},  \label{phifg} \\
\Psi_{\mu\nu}\left( \mathbf{x},t\right) & =2\kappa\int G_{\Psi}\left(
x^{\mu};x^{\mu^{\prime}}\right) T_{\mu\nu}\left( t^{\prime},\mathbf{x}%
^{\prime}\right) d^{4}x^{\prime}.  \label{psigf}
\end{align}
\begin{widetext}
The retarded Green's functions that appear in the previous equations are
described by the expression%
\begin{equation}
G_{X}\left(  x^{\mu};x^{\prime\mu}\right)  =-\frac{1}{4\pi}\frac{1}{c}\frac
{1}{s}\delta\left(  \tau-\frac{s}{c}\right)  +\frac{1}{4\pi}\frac{1}{c}%
\frac{m_{X}}{\sqrt{\tau^{2}-\left(  \frac{s}{c}\right)  ^{2}}}J_{1}\left(
m_{X}c\sqrt{\tau^{2}-\left(  \frac{s}{c}\right)  ^{2}}\right)  \Theta\left(
\tau-\frac{s}{c}\right)  , \label{Green}%
\end{equation}
where $\tau=t-t^{^{\prime}}$, $s=\left\vert \mathbf{s}\right\vert
=\left\vert \mathbf{x}-\mathbf{x}^{\prime}\right\vert $, and $X$ represents $%
\Phi$ or $\Psi$.
The function $\Theta$ is the Heaviside step function and $%
J_{1}$ is the Bessel function of the first kind. Details on the deduction of
Eq. (\ref{Green}) can be seen in Appendix A of Ref. \cite%
{PhysRevD.104.084061}.
\end{widetext}
\begin{widetext}
Substituting Green's function in Eqs. (\ref{phifg}) and (\ref{psigf}) and
considering large distances from the source, i.e. $s\simeq\left\vert
\mathbf{x}\right\vert \equiv r$, we get
\begin{equation}
\Phi\left(  \mathbf{x},t\right)  =\frac{\kappa}{12\pi}\frac{1}{r}\int T\left(
\mathbf{x}^{\prime},t_{r}\right)  d^{3}\mathbf{x}^{\prime}-\frac{m_{\Phi}%
}{12\pi}\kappa\int d^{3}\mathbf{x}^{\prime}\int_{\frac{r}{c}}^{\infty}d\bar
{t}\frac{J_{1}\left(  m_{\Phi}c\sqrt{\bar{t}^{2}-\left(  \frac{r}{c}\right)
^{2}}\right)  }{\sqrt{\bar{t}^{2}-\left(  \frac{r}{c}\right)  ^{2}}}T\left(
\mathbf{x}^{\prime},t-\bar{t}\right)  \label{Phi formal}%
\end{equation}
and%
\begin{equation}
\Psi_{ij}\left(  \mathbf{x},t\right)  =-\frac{\kappa}{2\pi r}\int
d^{3}\mathbf{x}^{\prime}T_{ij}\left(  \mathbf{x}^{\prime},t_{r}\right)
+\frac{m_{\Psi}\kappa}{2\pi}\int d^{3}\mathbf{x}^{\prime}\int_{\frac{r}{c}%
}^{\infty}d\bar{t}\frac{J_{1}\left(  m_{\Psi}c\sqrt{\bar{t}^{2}-\left(
\frac{r}{c}\right)  ^{2}}\right)  }{\sqrt{\bar{t}^{2}-\left(  \frac{r}%
{c}\right)  ^{2}}}T_{ij}\left(  \mathbf{x}^{\prime},t-\bar{t}\right)  .
\label{Psi formal}%
\end{equation}
\end{widetext}
Note that by the degrees of freedom of $\Psi_{\mu\nu}$ it is enough to
calculate its spatial components. See Sec. \ref{sec - DoF}.

The next step is to perform the multipolar expansion for the fields $\tilde {%
h}_{\mu\nu}$, $\Phi$ and $\Psi_{ij}$. The multipolar expansion for the
massless spin-$2$ field is well known from GR. The expansion procedure for
the scalar field was deduced in Ref. \cite{PhysRevD.104.084061} and results
in%
\begin{equation*}
\Phi\left( \mathbf{x},t\right) =\Phi^{M}\left( \mathbf{x},t\right)
+\Phi^{D}\left( \mathbf{x},t\right) +\Phi^{Q}\left( \mathbf{x},t\right)
+...,
\end{equation*}
where the monopole ($M$), dipole ($D$), and quadrupole ($Q$) contributions
are given by
\begin{widetext}
\begin{align}
\Phi^{M}\left(  \mathbf{x},t\right)   &  =\frac{\kappa}{12\pi r}\left[
c^{2}\mathcal{M}\left(  t_{r}\right)  \right]  -\frac{m_{\Phi}}{12\pi}%
\kappa\int_{0}^{\infty}d\bar{t}_{r}F_{\Phi}\left(  \bar{t}_{r}\right)  \left[
c^{2}\mathcal{M}\left(  \zeta\right)  \right]  ,\label{Phimon}\\
\Phi^{D}\left(  \mathbf{x},t\right)   &  =\frac{\kappa}{12\pi r}\left[
cn_{i}\left.  \frac{\partial\mathcal{M}^{i}}{\partial t}\right\vert _{t_{r}%
}\right]  -\frac{m_{\Phi}}{12\pi}\kappa\int_{0}^{\infty}d\bar{t}_{r}F_{\Phi
}\left(  \bar{t}_{r}\right)  \left[  cn_{i}\left.  \frac{\partial
\mathcal{M}^{i}}{\partial t}\right\vert _{\zeta}\right]  ,\label{Phidp}\\
\Phi^{Q}\left(  \mathbf{x},t\right)   &  =\frac{\kappa}{12\pi r}\left[
\frac{1}{2}n_{i}n_{j}\left.  \frac{\partial^{2}\mathcal{M}^{ij}}{\partial
t^{2}}\right\vert _{t_{r}}\right]  -\frac{m_{\Phi}}{12\pi}\kappa\int
_{0}^{\infty}d\bar{t}_{r}F_{\Phi}\left(  \bar{t}_{r}\right)  \left[  \frac
{1}{2}n_{i}n_{j}\left.  \frac{\partial^{2}\mathcal{M}^{ij}}{\partial t^{2}%
}\right\vert _{\zeta}\right]  , \label{Phiquad}%
\end{align}
\end{widetext}
with the unit vector $n^{i}$ pointing along the $x^{i}$ direction,
$\zeta=t_{r}-\bar{t}_{r}$ and%
\begin{equation}
F_{\Phi}\left(  \bar{t}_{r}\right)  =\frac{J_{1}\left(  m_{\Phi}c\sqrt
{2\bar{t}_{r}}\sqrt{\frac{\bar{t}_{r}}{2}+\frac{r}{c}}\right)  }{\sqrt
{2\bar{t}_{r}}\sqrt{\frac{\bar{t}_{r}}{2}+\frac{r}{c}}}. \label{F}%
\end{equation}
The mass moments built with the trace of the energy-momentum tensor are
defined as%
\begin{align}
\mathcal{M}\left( t\right) & \equiv\frac{1}{c^{2}}\int d^{3}\mathbf{x}%
^{\prime}T\left( \mathbf{x}^{\prime},t\right) ,  \label{MmonoSca} \\
\mathcal{M}^{i}\left( t\right) & \equiv\frac{1}{c^{2}}\int d^{3}\mathbf{x}%
^{\prime}T\left( \mathbf{x}^{\prime},t\right) x^{\prime i},  \label{MdipSca}
\\
\mathcal{M}^{ij}\left( t\right) & \equiv\frac{1}{c^{2}}\int d^{3}\mathbf{x}%
^{\prime}T\left( \mathbf{x}^{\prime},t\right) x^{\prime i}x^{\prime j}.
\label{MquaSca}
\end{align}
\begin{widetext}
The multipolar expansion for the massive spin-$2$ field follows similar
steps to those performed for the $\Phi $ field.
Thus, we get%
\[
\Psi _{ij}\left( \mathbf{x},t\right) =\frac{\kappa }{2\pi }\left[ -\frac{1}{r%
}S_{ij}\left( t_{r}\right) +m_{\Psi }\int_{0}^{\infty }d\bar{t}_{r}F_{\Psi
}\left( \bar{t}_{r}\right) S_{ij}\left( \zeta \right) -\frac{1}{r}\frac{n_{k}%
}{c}\left. \frac{\partial S^{ij,k}}{\partial t}\right\vert _{t_{r}}+m_{\Psi
}\int_{0}^{\infty }d\bar{t}_{r}F_{\Psi }\left( \bar{t}_{r}\right) \frac{n_{k}%
}{c}\left. \frac{\partial S^{ij,k}}{\partial t}\right\vert _{\zeta }+...%
\right] ,
\]%
\end{widetext}
where $F_{\Psi }\left( \bar{t}_{r}\right) $ is given by Eq. (\ref{F})
switching $m_{\Phi }\rightarrow m_{\Psi }$.The first two moments of the
stress tensor $T^{ij}$ are defined as
\begin{equation*}
S^{ij}\left( t\right) \equiv \int d^{3}\mathbf{x}^{\prime }T^{ij}\left(
\mathbf{x}^{\prime },t\right) \text{ \ }
\end{equation*}%
and

\begin{equation*}
S^{ij,k}\left( t\right) \equiv \int d^{3}\mathbf{x}^{\prime }T^{ij}\left(
\mathbf{x}^{\prime },t\right) x^{k}.
\end{equation*}%
Furthermore, the first moment $S^{ij}$ can be written as $S^{ij}=\frac{1}{2}%
\ddot{M}^{ij}$, where%
\begin{equation}
M^{ij}=\frac{1}{c^{2}}\int d^{3}\mathbf{x}^{\prime }T^{00}\left( \mathbf{x}%
^{\prime },t\right) x^{\prime i}x^{\prime j}  \label{Mom M qua}
\end{equation}%
is the usual quadrupole mass moment of general relativity.

Therefore, using the traceless-transverse ($TT$) gauge for $\tilde{h}_{\mu
\nu }$ and taking into account only the dominant terms of each mode, we have
\begin{align}
\left[ \tilde{h}_{ij}^{TT}\left( \mathbf{x},t\right) \right] _{Q}& =\frac{1}{%
r}\frac{\kappa }{4\pi }\Lambda _{ij,kl}\left( \mathbf{\hat{n}}\right) \ddot{M%
}^{kl}\left( t_{r}\right) ,  \label{spin2 massless} \\
\left[ \Psi _{ij}\left( \mathbf{x},t\right) \right] _{Q}& =-\frac{1}{r}\frac{%
\kappa }{4\pi }\ddot{M}_{ij}\left( t_{r}\right)  \label{spin2massive} \\
& +\frac{\kappa }{4\pi }m_{\Psi }\int_{0}^{\infty }d\bar{t}_{r}F_{\Psi
}\left( \bar{t}_{r}\right) \ddot{M}_{ij}\left( \zeta \right) ,  \notag \\
\Phi \left( \mathbf{x},t\right) & =\Phi ^{M}\left( \mathbf{x},t\right) +\Phi
^{D}\left( \mathbf{x},t\right) +\Phi ^{Q}\left( \mathbf{x},t\right) ,
\label{spin0}
\end{align}%
where $\Lambda _{ij,kl}$ is a projection tensor that selects the $TT$ gauge
\cite{magiorrebook}. We will see in the next section that for a binary
system with nonrelativistic dynamics $\Phi ^{M}=\Phi ^{D}=0$, and thus the
dominant term of the scalar part will also be a quadrupole term.

\section{Binary system in circular orbit\label{sec binary}}

We begin by writing the energy-momentum tensor for a nonrelativistic binary
point-mass system $m_{A}$ in the center-of-mass frame:
\begin{equation*}
T_{\mu \nu }=\sum_{A=1}^{2}m_{A}c^{2}\delta _{\mu }^{0}\delta _{\nu
}^{0}\delta ^{\left( 3\right) }\left( \mathbf{x}-\mathbf{x}_{A}\left(
t\right) \right) ,
\end{equation*}%
where $\mathbf{x}_{A}\left( t\right) $ is the vector representing the
trajectory of particle $A$. From this energy-momentum tensor and its trace,
we can calculate the mass moments for spin-$0$ and spin-$2$ modes in the
center-of-mass ($c.m.$) frame. So, by Eqs. (\ref{MmonoSca}), (\ref{MdipSca}), (%
\ref{MquaSca}), and (\ref{Mom M qua}) we get
\begin{eqnarray}
\mathcal{M}\left( t\right) &=&-m\text{, \ }\mathcal{M}^{i}\left( t\right) =0%
\text{ \ and \ }  \label{momentos} \\
\mathcal{M}^{ij}\left( t\right) &=&-\mu x_{0}^{i}\left( t\right)
x_{0}^{j}\left( t\right) =-M^{ij}\left( t\right) ,  \notag
\end{eqnarray}%
where $m=m_{1}+m_{2}$ is the total mass, $\mu =m_{1}m_{2}/m$ is the reduced
mass and $x_{0}^{i}\left( t\right) $ is the relative coordinate $\mathbf{x}%
_{0}=\mathbf{x}_{1}-\mathbf{x}_{2}$. The above expression shows that the
monopole and dipole contributions to the spin-$0$ mode are zero. This result
reflects the conservation of mass and linear momentum of a nonrelativistic
binary system.

The next step is to determine the trajectory $x_{0}^{i}\left( t\right) $.
For simplicity, we will consider a circular orbit of radius $R$ and angular
frequency $\omega_{s}$ positioned along the $\mathrm{XY}$-plane. In this
case, the relative coordinate is given by \footnote{For certain configurations of the binary system, the circular orbit
approximation can be quite realistic if we assume that the process of
circularization of elliptical orbits of general relativity also occurs in
quadratic gravity. The study of circularization in quadratic gravitation
will be carried out in a future work.}%
\begin{equation}
x_{0}^{i}\left( t\right) =\left( R\cos\left( \omega_{s}t+\frac{\pi}{2}%
\right) ,R\sin\left( \omega_{s}t+\frac{\pi}{2}\right) ,0\right) ,
\label{trajetoria}
\end{equation}
and the unit vector $n_{CM}^{i}=\left( 0,0,1\right) $. Note that as the
orbit is restricted to the $\mathrm{XY}$-plane, the mass moments $%
M^{13}=M^{23}=M^{33}=0$, which implies $\Psi_{B}=\Psi_{C}=\Psi_{D}=0$. Thus,
of the $5$ degrees of freedom associated with $\Psi_{ij}$, only the
transverse modes $\Psi_{+}$ and $\Psi_{+}$ are produced.

By fixing the coordinate system at the observer's point of view, we can
decompose the unit vector $n^{i}$ in terms of the polar angle $\theta$ and
the azimuthal angle $\phi$ as%
\begin{equation*}
n^{i}=\left( \sin\theta\sin\phi,\sin\theta\cos\phi,\cos\theta\right) .
\end{equation*}
In this reference frame, $\phi$ represents a phase in the $\mathrm{XY}$%
-plane, and $\theta$ is the angle between the normal of the orbit plane and
the line of sight. For more details, see Figs. $3.2$ and $3.6$ of Ref.
\cite{magiorrebook}.

The spin-$0$, massive spin-$2$, and massless spin-$2$ modes are obtained by
substituting the moments $M^{ij}$ and $\mathcal{M}^{ij}$ present in (\ref%
{momentos}) in Eqs. (\ref{spin2 massless}), (\ref{spin2massive}), (\ref%
{spin0}), and (\ref{Phiquad}), and then calculating the integrals that
contain the functions $F_{\Psi}$ and $F_{\Phi}$. The calculation of these
integrals is not trivial and can be found in Appendix $B$ of Ref. \cite%
{PhysRevD.104.084061}. Carrying out all these calculations, considering the
chirp mass $M_{c}\equiv m^{\frac{2}{5}}\mu^{\frac{3}{5}}$ and the Kepler's
third law $\omega_{s}^{2}=Gm/R^{3}$, we obtain%
\begin{widetext}
\begin{align}
\tilde{h}_{+}\left(  \mathbf{x},t\right)   &  =\frac{4c}{r}\left(
\frac{GM_{c}}{c^{3}}\right)  ^{\frac{5}{3}}\omega_{s}^{2/3}\left(
\frac{1+\cos^{2}\theta}{2}\right)  \cos\left[  2\omega_{s}\left(  t-\frac
{r}{c}\right)  +2\phi\right]  ,\label{h plus}\\
\tilde{h}_{\times}\left(  \mathbf{x},t\right)   &  =\frac{4c}{r}\left(
\frac{GM_{c}}{c^{3}}\right)  ^{\frac{5}{3}}\omega_{s}^{2/3}\cos\theta
\sin\left[  2\omega_{s}\left(  t-\frac{r}{c}\right)  +2\phi\right]  ,
\label{h cross}%
\end{align}
for massless tensorial modes,%
\begin{align}
\Psi_{+}\left(  \mathbf{x},t\right)   &  =\left\{
\begin{array}
[c]{c}%
-\frac{4c}{r}\left(  \frac{GM_{c}}{c^{3}}\right)  ^{\frac{5}{3}}\omega
_{s}^{2/3}\left(  \frac{1+\cos^{2}\theta}{2}\right)  \cos\left(  2\omega
_{s}t+2\phi\right)  \exp\left[  -m_{\Psi}r\sqrt{1-\left(  \frac{2\omega_{s}%
}{m_{\Psi}c}\right)  ^{2}}\right]  ,\text{ \ \ }2\omega_{s}<m_{\Psi}c\\
-\frac{4c}{r}\left(  \frac{GM_{c}}{c^{3}}\right)  ^{\frac{5}{3}}\omega
_{s}^{2/3}\left(  \frac{1+\cos^{2}\theta}{2}\right)  \cos\left[  2\omega
_{s}\left(  t-\left(  \frac{r}{c}\right)  \sqrt{1-\left(  \frac{m_{\Psi}%
c}{2\omega_{s}}\right)  ^{2}}\right)  +2\phi\right]  ,\text{ \ \ \ }%
2\omega_{s}>m_{\Psi}c
\end{array}
\right.  ,\label{Psi plus}\\
\Psi_{\times}\left(  \mathbf{x},t\right)   &  =\left\{
\begin{array}
[c]{c}%
-\frac{4c}{r}\left(  \frac{GM_{c}}{c^{3}}\right)  ^{\frac{5}{3}}\omega
_{s}^{2/3}\cos\theta\sin\left(  2\omega_{s}t+2\phi\right)  \exp\left[
-m_{\Psi}r\sqrt{1-\left(  \frac{2\omega_{s}}{m_{\Psi}c}\right)  ^{2}}\right]
,\text{ \ \ }2\omega_{s}<m_{\Psi}c\\
-\frac{4c}{r}\left(  \frac{GM_{c}}{c^{3}}\right)  ^{\frac{5}{3}}\omega
_{s}^{2/3}\cos\theta\sin\left[  2\omega_{s}\left(  t-\left(  \frac{r}%
{c}\right)  \sqrt{1-\left(  \frac{m_{\Psi}c}{2\omega_{s}}\right)  ^{2}%
}\right)  +2\phi\right]  ,\text{ \ \ \ }2\omega_{s}>m_{\Psi}c
\end{array}
\right.  , \label{Psi cross}%
\end{align}
for the massive tensorial modes and
\begin{equation}
\Phi\left(  \mathbf{x},t\right)  =\left\{
\begin{array}
[c]{c}%
\frac{2c}{3r}\left(  \frac{GM_{c}}{c^{3}}\right)  ^{\frac{5}{3}}\omega
_{s}^{2/3}\sin^{2}\theta\cos\left(  2\omega_{s}t+2\phi\right)  \exp\left[
-m_{\Phi}r\sqrt{1-\left(  \frac{2\omega_{s}}{m_{\Phi}c}\right)  ^{2}}\right]
,\text{ \ \ }2\omega_{s}<m_{\Phi}c\\
\frac{2c}{3r}\left(  \frac{GM_{c}}{c^{3}}\right)  ^{\frac{5}{3}}\omega
_{s}^{2/3}\sin^{2}\theta\cos\left[  2\omega_{s}\left(  t-\left(  \frac{r}%
{c}\right)  \sqrt{1-\left(  \frac{m_{\Phi}c}{2\omega_{s}}\right)  ^{2}%
}\right)  +2\phi\right]  ,\text{ \ \ \ }2\omega_{s}>m_{\Phi}c
\end{array}
\right.  , \label{Phi}%
\end{equation}
for massive scalar mode.
\end{widetext}
The most relevant point of the solutions above is that the massive modes
have two distinct regimes. The first of these is a damping regime that
occurs when\ $2\omega_{s}<m_{X}c$.\footnote{%
Remembering that $X$ represents $\Phi$ or $\Psi$.} In this regime we do not
have a wave solution, and the massive modes only contribute with a temporal
modulation for the gravitational field that exponentially decays when moving
away from the source. The second regime, called the oscillatory regime,
occurs when $2\omega_{s}>m_{X}c$. It is only in this regime that the source
emits gravitational waves associated with the massive modes.

The solutions for $\tilde{h}_{+},\Psi_{+}$\ and $\tilde{h}%
_{\times},\Psi_{\times}$\ in the oscillatory regime are waves of the same
amplitude, same frequency and different wave numbers, which provides an
interpretation of interference effects. In this sense, it is convenient to
introduce
\begin{equation}
\Theta_{+,\times}=\tilde{h}_{+,\times}+\Psi_{+,\times}.
\label{Completespin2}
\end{equation}
Equation (\ref{Completespin2}) allows us to interpret the spin-$2$ waves as
a single structure generated by interference between the fields $\tilde {h}%
_{+,\times}$ and $\Psi_{+,\times}$. Note that due to the difference in sign
between $\tilde{h}_{+,\times}$\ and $\Psi_{+,\times}$,\ what we have is a
destructive interference effect. Furthermore, at the limit of $m_{\Psi
}\rightarrow0$,\ this destructive interference is complete, and in this
case, the emission of tensorial modes does not occur. We will see in the
next sections that physical interpretations of the binary system are
consistent only when we treat massive and massless spin-$2$ modes together.

\subsection{Gravitational energy-momentum tensor and the power radiated}
The low-frequency effects of second-order terms contribute to background
changes \cite{PhysRev.166.1272}. These terms, obtained from space-time
averages $\left\langle ...\right\rangle $, generate the gravitational
energy-momentum tensor $t_{\mu \alpha }$ given by
\begin{equation}
t_{\mu \alpha }=-\frac{c^{4}}{8\pi G}\left[ \left\langle G_{\mu \alpha
}^{\left( 2\right) }\right\rangle +\gamma \left\langle H_{\mu \alpha
}^{\left( 2\right) }\right\rangle -2\alpha \left\langle I_{\mu \alpha
}^{\left( 2\right) }\right\rangle \right] ,  \label{pseudo tensor def}
\end{equation}%
\begin{widetext}
where%
\begin{align}
G_{\mu\alpha}^{\left(  2\right)  }  &  =R_{\mu\alpha}^{\left(  2\right)
}-\frac{1}{2}\eta_{\mu\alpha}R^{\left(  2\right)  }-\frac{1}{2}h_{\mu\alpha
}R^{\left(  1\right)  },\label{Gmunu2}\\
H_{\mu\alpha}^{\left(  2\right)  }  &  =R^{\left(  1\right)  }R_{\mu\alpha
}^{\left(  1\right)  }-\frac{1}{4}R^{\left(  1\right)  }R^{\left(  1\right)
}\eta_{\mu\alpha}+h_{\mu\alpha}\partial^{\sigma}\partial_{\sigma}R^{\left(
1\right)  }+\eta_{\mu\alpha}\square^{\left(  1\right)  }R^{\left(  1\right)
}+\eta_{\mu\alpha}\partial^{\sigma}\partial_{\sigma}R^{\left(  2\right)
}-\nabla_{\mu}^{\left(  1\right)  }\partial_{\alpha}R^{\left(  1\right)
}-\partial_{\mu}\partial_{\alpha}R^{\left(  2\right)  },\label{Hmunu2}\\
I_{\mu\alpha}^{\left(  2\right)  }  &  =\left(  \nabla^{\nu}\nabla^{\beta
}\right)  ^{\left(  1\right)  }C_{\mu\nu\alpha\beta}^{\left(  1\right)
}+\partial^{\nu}\partial^{\beta}C_{\mu\nu\alpha\beta}^{\left(  2\right)
}+\frac{1}{2}R^{\nu\beta\left(  1\right)  }C_{\mu\nu\alpha\beta}^{\left(
1\right)  }. \label{Imunu2}%
\end{align}
After a long calculation presented in Appendix \ref{sec - Ap GraEneMoTen}%
, we get%
\begin{equation}
t_{\mu\alpha}=\frac{c^{4}}{8\pi G}\left[ \frac{1}{4}\left\langle
\partial_{\mu}\tilde{h}_{\nu\beta}\partial_{\alpha}\tilde{h}^{\nu\beta
}\right\rangle -\frac{1}{4}\left\langle
\partial_{\mu}\Psi_{\nu\beta}\partial_{\alpha}\Psi^{\nu\beta}\right\rangle +%
\frac{3}{2}\left\langle \partial_{\alpha}\Phi\partial_{\mu}\Phi\right\rangle
+\frac{1}{2}\left\langle \partial_{\mu}\tilde{h}_{\nu\beta}\partial_{\alpha}%
\Psi^{\nu\beta }\right\rangle \right] .   \label{t_munu}
\end{equation}
\end{widetext}An important point to be discussed is the opposite (negative)
sign that appears in the second term of Eq. (\ref{t_munu}) \cite{massive}. Although
potentially pathological, this negative sign is expected because the
Weyl-Weyl term presents Ostrogradsky instability, implying that the
Hamiltonian density is not positive definite.\footnote{%
This phenomenon has been known for a long time in the context of quadratic
gravity quantization \cite{PhysRevD.16.953}.} However, in the context we are
studying, it is possible to avoid any pathology by considering the massive
and massless spin-$2$ modes as a single structure defined in Eq. (\ref%
{Completespin2}). In this context, neglecting the scalar mode, we have two
different situations:

\begin{itemize}
\item For $2\omega_{s}<m_{\Psi}c$ (damping regime), the massive spin-$2$
part does not emit radiation -- $\left\langle
\partial_{\mu}\Psi_{\nu\beta}\partial_{\alpha}\Psi^{\nu\beta}\right\rangle
=0 $ -- and the system behaves as in general relativity.

\item For $2\omega_{s}>m_{\Psi}c$ (oscillatory regime), spin-$2$ modes emit
two waves that interfere destructively in such a way that the radiated
energy is always less than the case of pure GR but always positive, i.e.,$%
\left\langle \partial_{\mu}\tilde{h}_{\nu\beta}\partial_{\alpha}\tilde{h}%
^{\nu\beta}\right\rangle -\left\langle \partial_{\mu}\Psi_{\nu\beta
}\partial_{\alpha}\Psi^{\nu\beta}\right\rangle >0$.
\end{itemize}

The radiated power per solid angle unit is given by%
\begin{equation}
\frac{dP}{d\Omega}=-cr^{2}t_{01}.  \label{Pot ang}
\end{equation}
Due to the functional form of $\tilde{h}_{+.\times}$ their radial
derivatives $\partial_{1}$ can be switched to time derivatives $\partial_{0}$
up to $\mathcal{O}\left( 1/r^{2}\right) $:
\begin{equation*}
\partial_{1}\tilde{h}_{+.\times}=-\partial_{0}\tilde{h}_{+.\times}.
\end{equation*}
\begin{widetext}
Furthermore, using analogous reasoning for $\Psi_{+.\times}$ and $\Phi$, we
obtain%
\begin{align*}
\partial_{1}\Psi_{+}  &  =\left\{
\begin{array}
[c]{c}%
\sqrt{\left(  \frac{m_{\Psi}c}{2\omega_{s}}\right)  ^{2}-1}\cot\left(
2\omega_{s}t+2\phi\right)  \partial_{0}\Psi_{+},\text{ \ \ }2\omega
_{s}<m_{\Psi}c\\
-\sqrt{1-\left(  \frac{m_{\Psi}c}{2\omega_{s}}\right)  ^{2}}\partial_{0}%
\Psi_{+},\text{ \ \ \ }2\omega_{s}>m_{\Psi}c
\end{array}
\right.  ,\\
\partial_{1}\Psi_{\times}  &  =\left\{
\begin{array}
[c]{c}%
-\sqrt{\left(  \frac{m_{\Psi}c}{2\omega_{s}}\right)  ^{2}-1}\tan\left(
2\omega_{s}t+2\phi\right)  \partial_{0}\Psi_{\times},\text{ \ \ }2\omega
_{s}<m_{\Psi}c\\
-\sqrt{1-\left(  \frac{m_{\Psi}c}{2\omega_{s}}\right)  ^{2}}\partial_{0}%
\Psi_{\times},\text{ \ \ \ }2\omega_{s}>m_{\Psi}c
\end{array}
\right.
\end{align*}
and%
\[
\partial_{1}\Phi=\left\{
\begin{array}
[c]{c}%
\sqrt{\left(  \frac{m_{\Phi}c}{2\omega_{s}}\right)  ^{2}-1}\cot\left(
2\omega_{s}t+2\phi\right)  \partial_{0}\Phi,\text{ \ \ }2\omega_{s}<m_{\Phi
}c\\
-\sqrt{1-\left(  \frac{m_{\Phi}c}{2\omega_{s}}\right)  ^{2}}\partial_{0}%
\Phi,\text{ \ \ \ }2\omega_{s}>m_{\Phi}c
\end{array}
\right.  .
\]
Substituting these last results in Eq. (\ref{Pot ang}) and calculating the
spatial-time averages, we get \cite{PhysRevD.104.084061}
\begin{align}
\frac{dP}{d\Omega}  & =-\frac{2G\mu^{2}R^{4}\omega_{s}^{6}}{\pi c^{5}}\left[
\left(  \left(  \frac{1+\cos^{2}\theta}{2}\right)  ^{2}+\cos^{2}\theta\right)
\left(  1-\Theta\left(  2\omega_{s}-m_{\Psi}c\right)  \sqrt{1-\left(
\frac{m_{\Psi}c}{2\omega_{s}}\right)  ^{2}}\right)  \right.
\label{Pot irr ang}\\
& \left.  +\Theta\left(  2\omega_{s}-m_{\Phi}c\right)  \sqrt{1-\left(
\frac{m_{\Phi}c}{2\omega_{s}}\right)  ^{2}}\frac{1}{12}\sin^{4}\theta\right]
.\nonumber
\end{align}
The Heaviside functions $\Theta\left( 2\omega_{s}-m_{\Psi}c\right) $ and $%
\Theta\left( 2\omega_{s}-m_{\Phi}c\right) $ indicate that only the
oscillatory regimes contribute to the radiated power. It is also worth
noting that the cross term $\left\langle \partial_{0}\tilde{h}%
_{\nu\beta}\partial _{1}\Psi^{\nu\beta}\right\rangle $ cancels out when
averaged in time and space.
One additional integration gives the expression for the total power radiated:%
\begin{equation}
P=-\frac{32}{5}\frac{c^{5}}{G}\left(  \frac{GM_{c}\omega_{s}}{c^{3}}\right)
^{\frac{10}{3}}\left[  1-\Theta\left(  2\omega_{s}-m_{\Psi}c\right)
\sqrt{1-\left(  \frac{m_{\Psi}c}{2\omega_{s}}\right)  ^{2}}+\frac
{\Theta\left(  2\omega_{s}-m_{\Phi}c\right)  }{18}\sqrt{1-\left(
\frac{m_{\Phi}c}{2\omega_{s}}\right)  ^{2}}\right]  . \label{Pot irr Tot}%
\end{equation}
\end{widetext}
Considering the tensorial part in the oscillatory regime and the scalar part
in the damping regime, we get%
\begin{equation*}
P_{spin2}=-\frac{32}{5}\frac{c^{5}}{G}\left( \frac{GM_{c}\omega_{s}}{c^{3}}%
\right) ^{\frac{10}{3}}\left[ 1-\sqrt{1-\left( \frac{m_{\Psi}c}{2\omega_{s}}%
\right) ^{2}}\right] .
\end{equation*}
This last expression corroborates the previous statement that the emitted
energy is always positive definite. In fact, for $m_{\Psi}c=2\omega_{s}$, we
are at the threshold of the damping regime and the system loses energy
exclusively by the massless spin-$2$ mode. As $m_{\Psi}$ decreases, the
destructive interference effect occurs between $\tilde{h}_{+,\times}$\ and $%
\Psi_{+,\times}$ and the energy loss decreases monotonically. In the
limiting case where $m_{\Psi}\rightarrow0$, the destructive interference is
maximum and the energy loss ceases to exist via tensorial modes. It is worth
remembering that the limit $m_{\Psi}\rightarrow0\Rightarrow\alpha\rightarrow%
\infty$ is nonphysical since the term $C_{\mu\nu\alpha\beta}C^{\mu\nu\alpha%
\beta}$ diverges in the action (\ref{acao0}).

\section{Inspiral phase of binary black holes\label{sec inspiral}}

Let us consider the case of a binary system consisting of two static black
holes with spherical symmetry. For simplicity, we will adopt the
Schwarzschild solution as a static solution, although, in quadratic gravity,
other solutions might exist \cite{PhysRevD.92.124019,PhysRevLett.114.171601}%
. Thus, in the weak field regime, the gravitational potential is reduced to
the Newtonian potential.

The next step is to establish the balance equation that determines the
energy loss of the system through the emission of gravitational waves:%
\begin{equation}
P=-\frac{dE_{orbit}}{dt},  \label{BalanceEqFun}
\end{equation}
where $P$ is the total radiated power and%
\begin{equation}
\frac{dE_{orbit}}{dt}=\frac{Gm_{1}m_{2}}{2R^{2}}\dot{R},  \label{dEorbit}
\end{equation}
is obtained in the approximation of quasicircular orbits \cite{magiorrebook}%
. Substituting Eqs. (\ref{Pot irr Tot}) and (\ref{dEorbit}) in the
expression (\ref{BalanceEqFun}) and using Kepler's third law, we get%
\begin{widetext}
\begin{equation}
\dot{\omega}_{s}=\frac{96}{5}\left(  \frac{GM_{c}}{c^{3}}\right)  ^{\frac
{5}{3}}\omega_{s}^{\frac{11}{3}}\left\{  1-\Theta\left(  2\omega_{s}-m_{\Psi
}c\right)  \sqrt{1-\left(  \frac{m_{\Psi}c}{2\omega_{s}}\right)  ^{2}}%
+\frac{\Theta\left(  2\omega_{s}-m_{\Phi}c\right)  }{18}\sqrt{1-\left(
\frac{m_{\Phi}c}{2\omega_{s}}\right)  ^{2}}\right\}  . \label{BalanceEq}%
\end{equation}
\end{widetext}
The equation (\ref{BalanceEq}) determines the variation of the orbital
frequency in the inspiral phase. The term within curly brackets contains the
usual contribution from GR (first term) and additional contributions from
the massive spin-$2$ and spin-$0$ modes. The massive modes only
contribute when $\Theta\left( 2\omega_{s}-m_{\Psi}c\right) =1$ or $%
\Theta\left( 2\omega_{s}-m_{\Phi}c\right) =1$, that is, only when the
solutions are in the oscillatory regime.

To analyze the effect of massive modes in the binary system, we will divide
the study into two cases: in the first one, we only consider the tensorial
mode in the oscillatory regime; in the second, we only take into account the
scalar mode.

\textbf{GR plus massive spin}-$2$\textbf{\ mode:} In this case we have $%
2\omega_{s}>m_{\Psi}c$ and $2\omega_{s}<m_{\Phi}c$, so that Eq. (%
\ref{BalanceEq}) contains only the first two terms. By integrating Eq. (\ref%
{BalanceEq}), we get the curves shown in Fig. \ref{fig1}:
\begin{figure}[h!]
\includegraphics[scale=0.5]{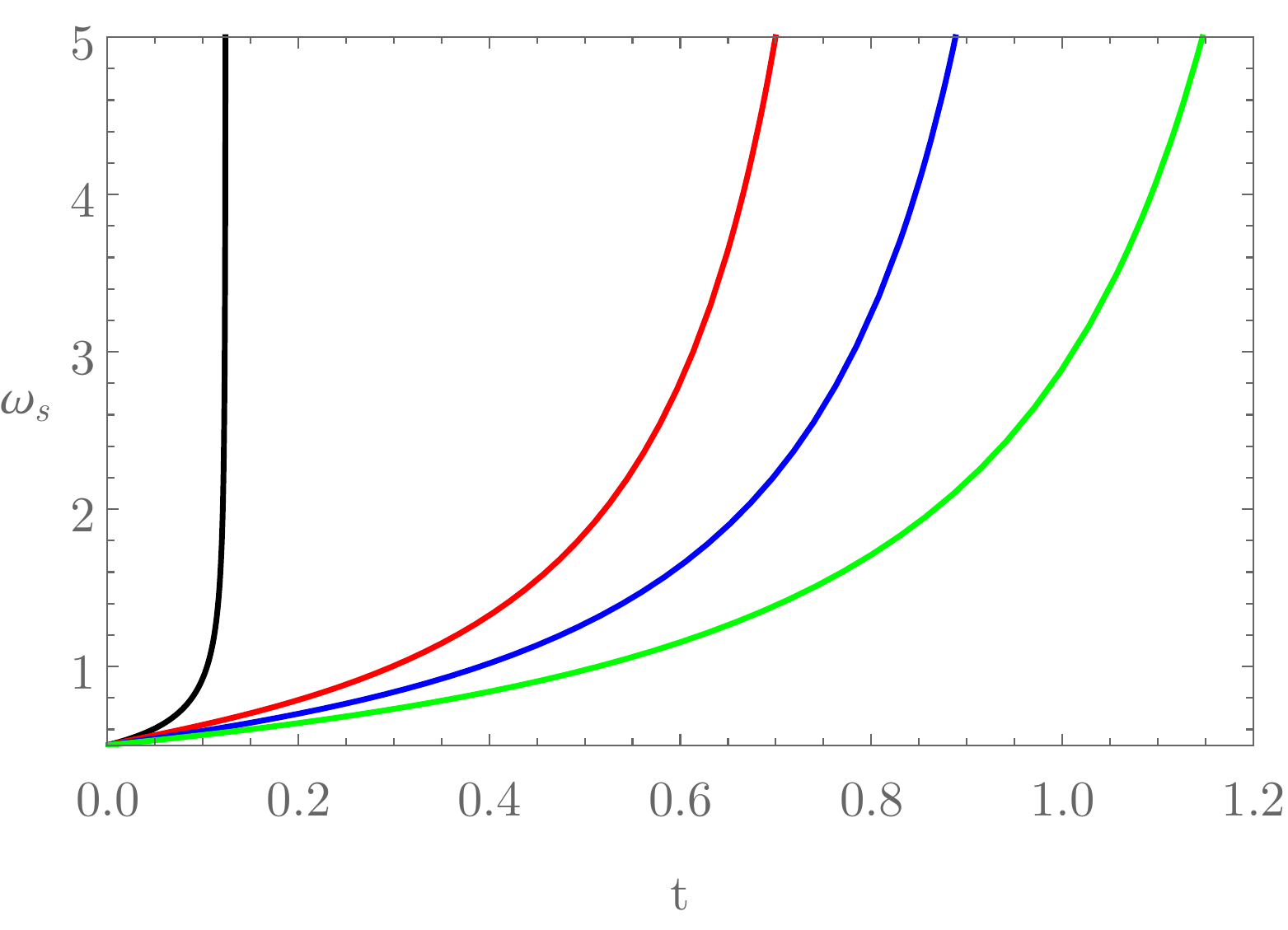}
\caption{Numerical solution of the balance equation in units of $%
GM_{c}/c^{3}=1$, with initial condition $\protect\omega\left( 0\right) =0.5$
and taking into account only spin-$2$ modes. The red, blue and green curves
are constructed with $m_{\Psi}c=1$, $m_{\Psi}c=0.9$ and $m_{\Psi}c=0.8$,
respectively. The black curve represents the pure GR solution.}
\label{fig1}
\end{figure}

Figure \ref{fig1} presents solutions for the combined spin-$2$ structure,
i.e. $\Theta_{\mu\nu}=\tilde{h}_{\mu\nu}+\Psi_{\mu\nu}$, and the pure case
of general relativity. This figure clearly shows that the coalescence time
increases as the value $m_{\Psi}c$ decreases. It occurs because the
reduction of $m_{\Psi}c$ makes the process of destructive interference
between the $\tilde{h}_{\mu\nu}$ and $\Psi_{\mu\nu}$ modes more effective,
and consequently, the binary system loses energy more slowly. It is
important to note that for $m_{\Psi}c>0$, i.e. $\alpha$ finite, coalescence
always occurs.

\begin{figure}[h!]
\includegraphics[scale=0.5]{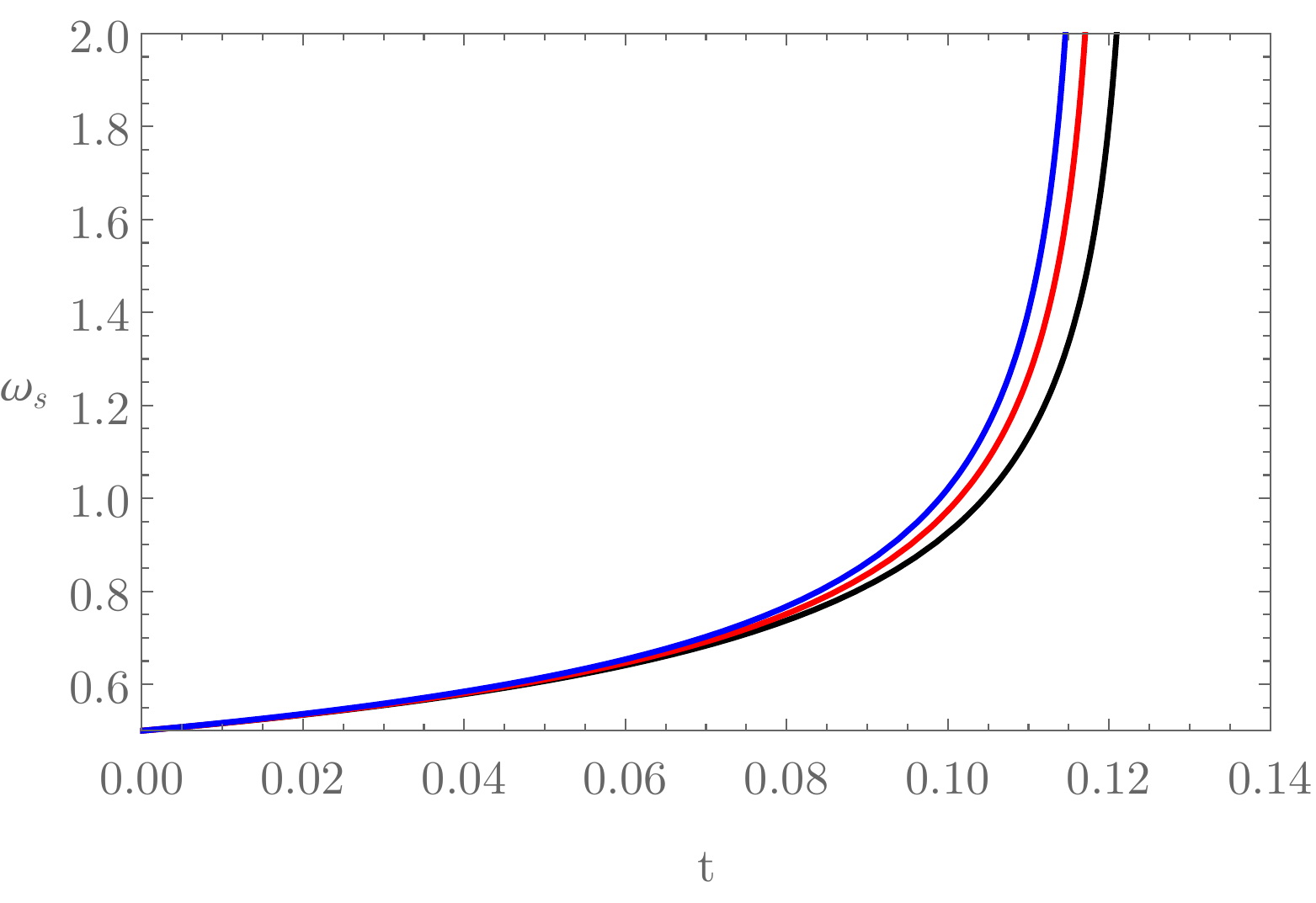}
\caption{Numerical solution of the balance equation in units of $%
GM_{c}/c^{3}=1$, with initial condition $\protect\omega\left( 0\right) =0.5$
and taking into account only the spin-$0$ and massless spin-$2$ modes. The
red and blue curves are constructed with $m_{\Phi}c=1$ and $m_{\Phi}c=0.1$,
respectively. The black curve represents the pure general relativity
solution.}
\label{fig2}
\end{figure}

\textbf{GR plus massive spin}-$0$\textbf{\ mode:} In this other
configuration we have $2\omega_{s}<m_{\Psi}c$ and $2\omega_{s}>m_{\Phi}c$,
and therefore only the first and third terms are present in equation (\ref%
{BalanceEq}). By integrating Eq. (\ref{BalanceEq}) in this context, we get
the results shown in Fig. \ref{fig2}.

Figure \ref{fig2} presents $\omega_{s}\left( t\right) $ in the inspiral
phase considering that the binary system loses energy through the $\tilde{h}%
_{\mu \nu}$ and $\Phi$ modes. From this figure, we can see that the lower
the value of $m_{\Phi}c$, the more effective is the energy loss via scalar
mode, and consequently, the coalescence occurs earlier. Furthermore, the
proximity of the curves in Fig. \ref{fig2} indicates that the scalar mode
carries considerably less energy than the tensorial one, i.e. the orbital
dynamics of the system is essentially determined by the $\tilde{h}_{\mu\nu}$
mode.

\section{Spin-2 waveform\label{sec spin 2}}

Once the orbital dynamics of the binary system is established, we will study
what is the waveform of the unique spin-$2$ structure detected at a certain
point in space.\footnote{%
During this section, we assume that there is no scalar mode emission.} We
saw earlier that the complete spin-$2$ wave is composed of two modes, namely
$\tilde{h}_{+.\times}$ and $\Psi_{+.\times}$, which interfere destructively.
At the time of emission, these two modes have the same frequency and
amplitude, but this is no longer true at the time of detection. The main
point of this analysis is that the massive and massless modes propagate at
different velocities. While the $\tilde{h}_{+.\times}$ mode propagates with
velocity $c$, the $\Psi_{+.\times}$ mode propagates with group velocity

\begin{equation*}
v_{g}\left( t\right) =\frac{d\omega}{dk},
\end{equation*}
where $\omega=c\sqrt{k^{2}+m_{\Psi}^{2}}$ is the dispersion relation of the
massive mode. Remembering that in the approximation of quasicircular orbits
we have $\omega=2\omega_{s}$, we can write the velocity $v_{g}$ as%
\begin{equation}
v_{g}=c\sqrt{1-\left( \frac{m_{\Psi}c}{2\omega_{s}}\right) ^{2}},
\label{v group}
\end{equation}
with $v_{g}<c$. The difference in propagation velocities between the two modes
generates the phenomenon of dispersion in the complete spin-$2$ wave.

The combination of massive and massless modes at detection time $t_{d}$
occurs with waves that were emitted at different times and therefore have
different frequencies. If the $\tilde{h}_{+.\times}$ mode was emitted in $%
t_{e}$, the $\Psi_{+.\times}$ mode was necessarily emitted earlier in $%
t_{e_{m}}$. In this case, the frequency $\omega_{s}\left( t_{e}\right) $
associated with the massless mode is greater than the frequency $%
\omega_{s}\left( t_{e_{m}}\right) $ associated with the massive mode.

Based on the previous discussion, we can write the relationships between the
detection time and the emission times as%
\begin{align}
t_{e} & =t_{d}-\frac{r}{c},  \label{retarded time e} \\
t_{e_{m}} & =t_{d}-\frac{r}{v_{g}\left( t_{e_{m}}\right) }.
\label{retarded time em}
\end{align}
Combining these two expressions and using Eq. (\ref{v group}), we get%
\begin{equation}
t_{e_{m}}=t_{e}+\frac{r}{c}\left( 1-\frac{1}{\sqrt{1-\left( \frac{m_{\Psi}c}{%
2\omega_{s}\left( t_{e_{m}}\right) }\right) ^{2}}}\right) .
\label{Eq algebrica tem}
\end{equation}
The previous equation is an algebraic equation for $t_{e_{m}}$ which
together with Eq. (\ref{BalanceEq}) allows us to determine $%
t_{e_{m}}$ and $\omega_{s}\left( t_{e_{m}}\right) $.

In order to exemplify the phenomenon of dispersion in the spin-$2$ complete
wave, we consider a hypothetical binary system of $M_{c}=10M_{\odot}$
located at a distance of $10Kpc$. We also assume $m_{\Psi}c=0.02$ $s^{-1}$
and neglect the scalar mode emission.\footnote{%
This can be obtained by considering $m_{\Phi}c$ much higher than the typical
orbital frequency of the binary system.} For convenience $t_{e}$ is adopted
as a time evolution variable and we consider $\omega_{s}\left(
t_{e}=0\right) =0.01$ $Hz$. This initial condition causes the $%
\Psi_{+.\times}$ mode to transition from the damping regime to the
oscillatory one exactly at $t_{e}=0$.

The next step is to numerically solve the balance equation (\ref{BalanceEq})
obtaining $\omega _{s}\left( t\right) $, and then determine $t_{e_{m}}$ by
Eq. (\ref{Eq algebrica tem}) considering different values of $t_{e}$. To
facilitate this analysis, we define the functions%
\begin{equation*}
f\left( t_{e_{m}}\right) =t_{e_{m}}
\end{equation*}%
and%
\begin{equation*}
g\left( t_{e_{m}}\right) =t_{e}+\frac{r}{c}\left( 1-\frac{1}{\sqrt{1-\left(
\frac{m_{\Psi }c}{2\omega _{s}\left( t_{e_{m}}\right) }\right) ^{2}}}\right)
.
\end{equation*}%

Figure \ref{fig3} shows plots of the functions $f\left( t_{e_{m}}\right) $
and $g\left( t_{e_{m}}\right) $ for three distinct values of $t_{e}$.
\begin{figure}[h!]
\includegraphics[scale=0.5]{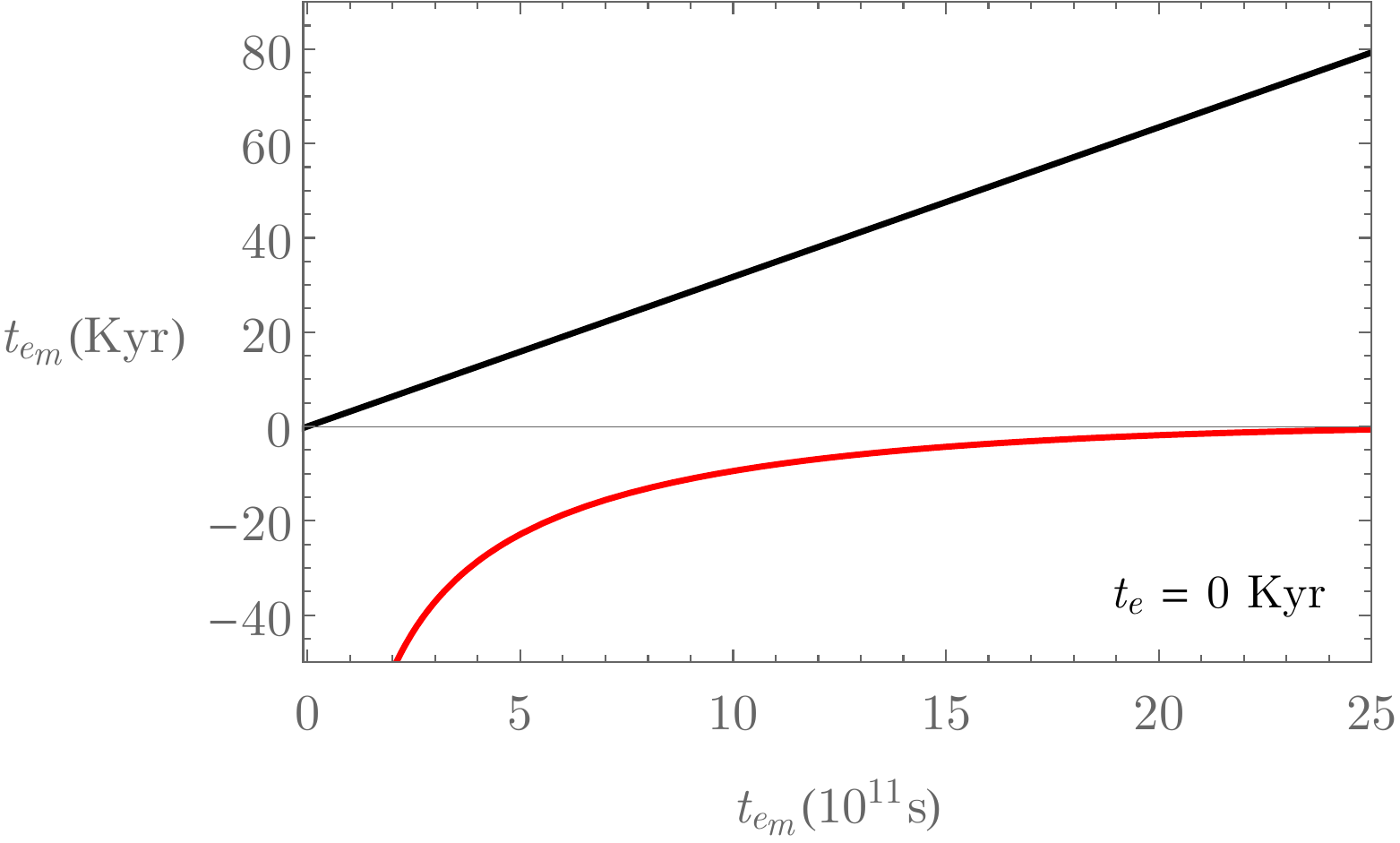}
\end{figure}
\begin{figure}[h!]
\includegraphics[scale=0.5]{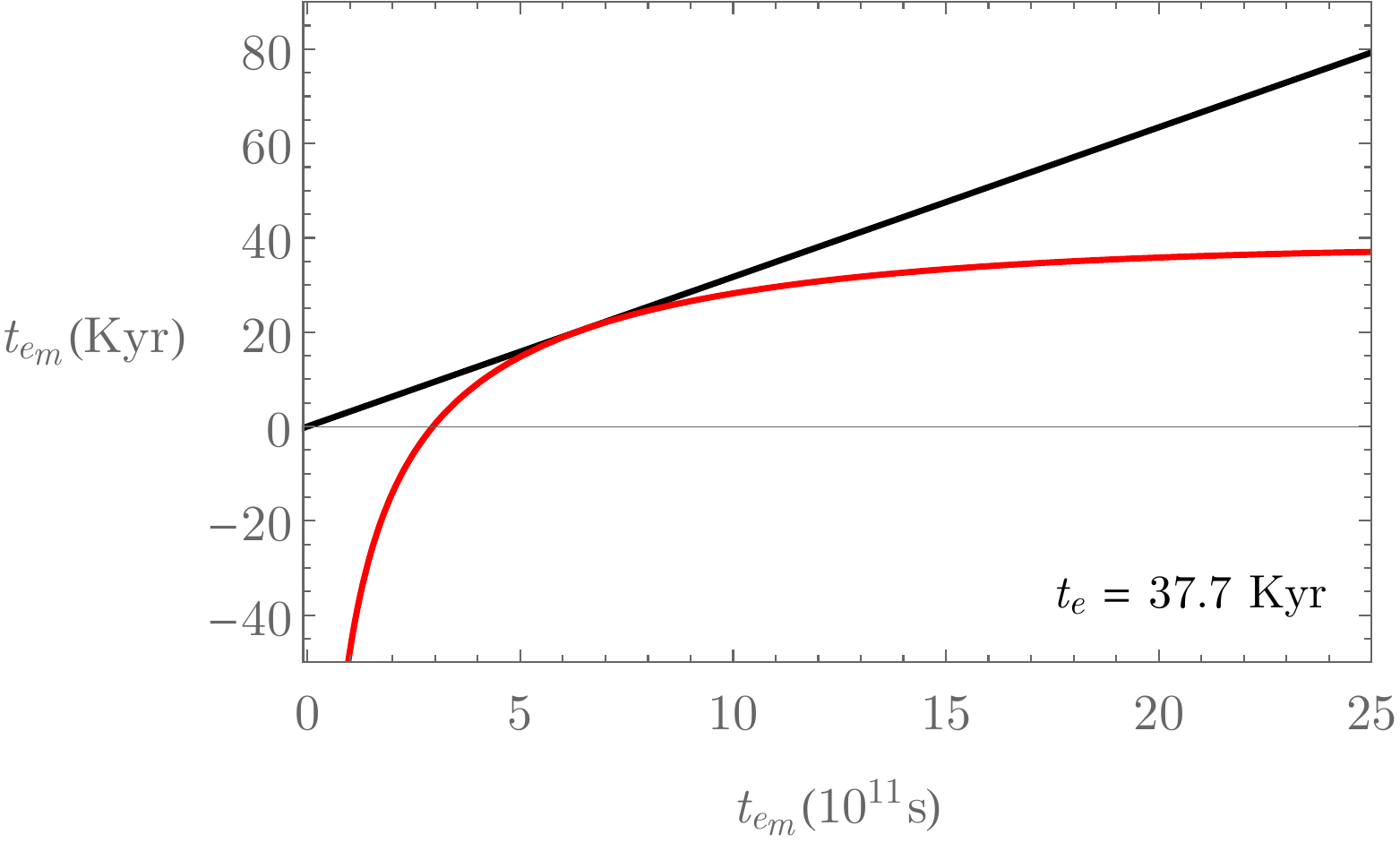}
\end{figure}
\begin{figure}[h!]
\includegraphics[scale=0.5]{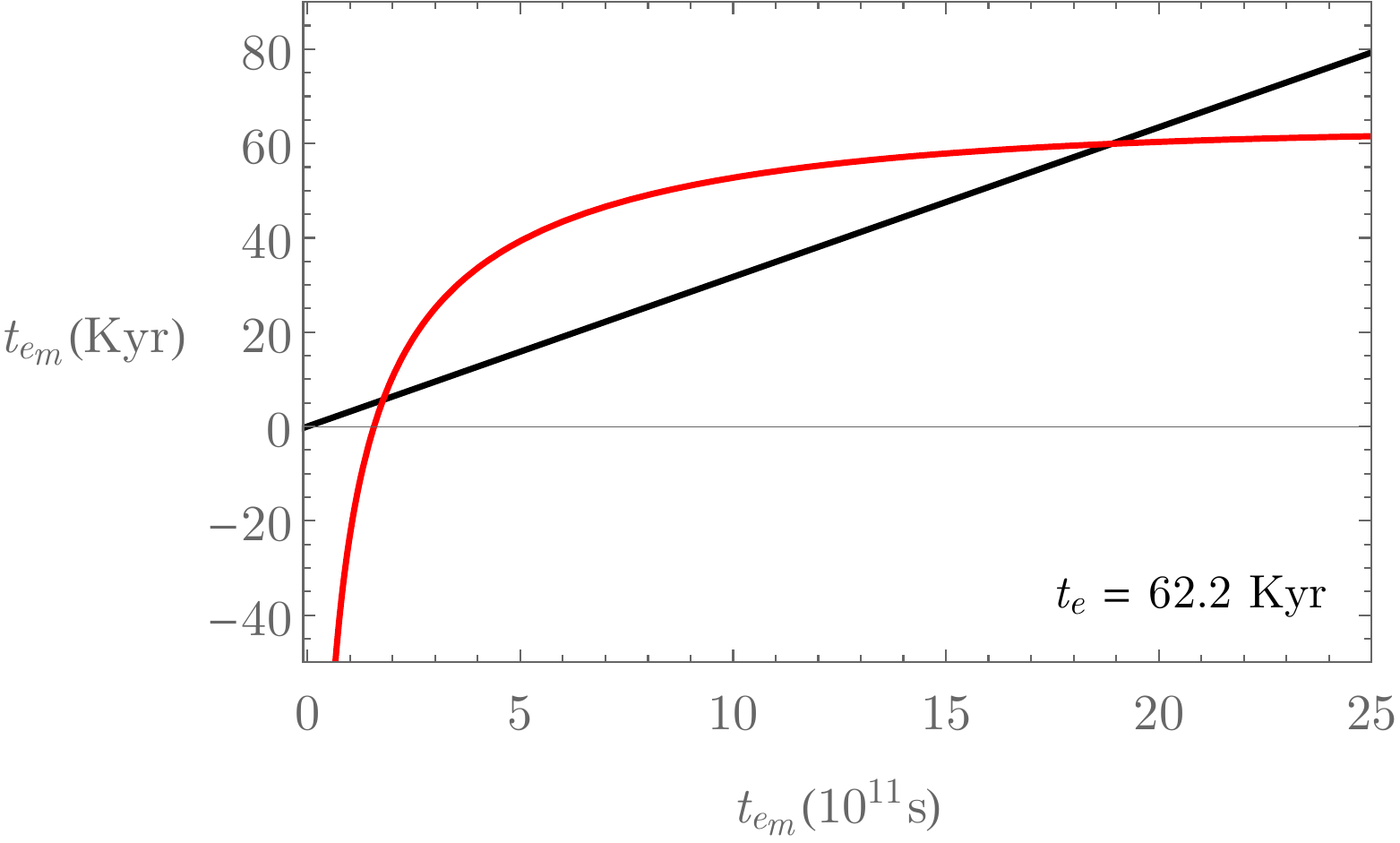}
\caption{Plots of the functions $f\left( t_{e_{m}}\right) $ (black) and $%
g\left( t_{e_{m}}\right) $ (red) considering three values of $t_{e}$. In the
construction of $g\left( t_{e_{m}}\right) $, we use the numerical solution
of Eq. (\protect\ref{BalanceEq}) with $M_{c}=10M_{\odot }$, $r=10$ $Kpc$, $%
m_{\Psi }c=0.02$ $s^{-1}$ and initial condition $\protect\omega _{s}\left(
0\right) =0.01$ $Hz$.}
\label{fig3}
\end{figure}

The plots sequence in Fig. \ref{fig3} shows that as $t_{e}$ increases, the
red curve shifts upwards. Furthermore, from $t_{e}=37.7$ $Kyr$, this curve
intercepts the black curve indicating the existence of solutions of Eq. (\ref%
{Eq algebrica tem}). Physically this means that the first wave fronts of the
massive modes take $37.7$ $Kyr$ to travel a distance of $10$ $Kpc$.
Furthermore, due to the nonlinearity of velocity $v_{g}\left( t\right) $, we
see that the complete spin-$2$ wave detected is composed by the
superposition of a massless mode $\tilde{h}_{+.\times}$ and two massive
modes $\Psi _{+.\times}^{\left( 1,2\right) }$. This fact is evidenced by the
double solution of Eq. (\ref{Eq algebrica tem}), which occurs from $%
t_{e}=37.7$ $Kyr$. Figure \ref{fig4} shows waveforms $\Theta_{+}=\tilde{h}%
_{+}\left( \omega _{s}^{e}\right) +\Psi_{+}^{\left( 1\right) }\left(
\omega_{s}^{e_{m\left( 1\right) }}\right) +\Psi_{+}^{\left( 2\right) }\left(
\omega _{s}^{e_{m\left( 2\right) }}\right) $ considering the same three
values of $t_{e}$ presented above.
\begin{figure}[h!]
\includegraphics[scale=0.5]{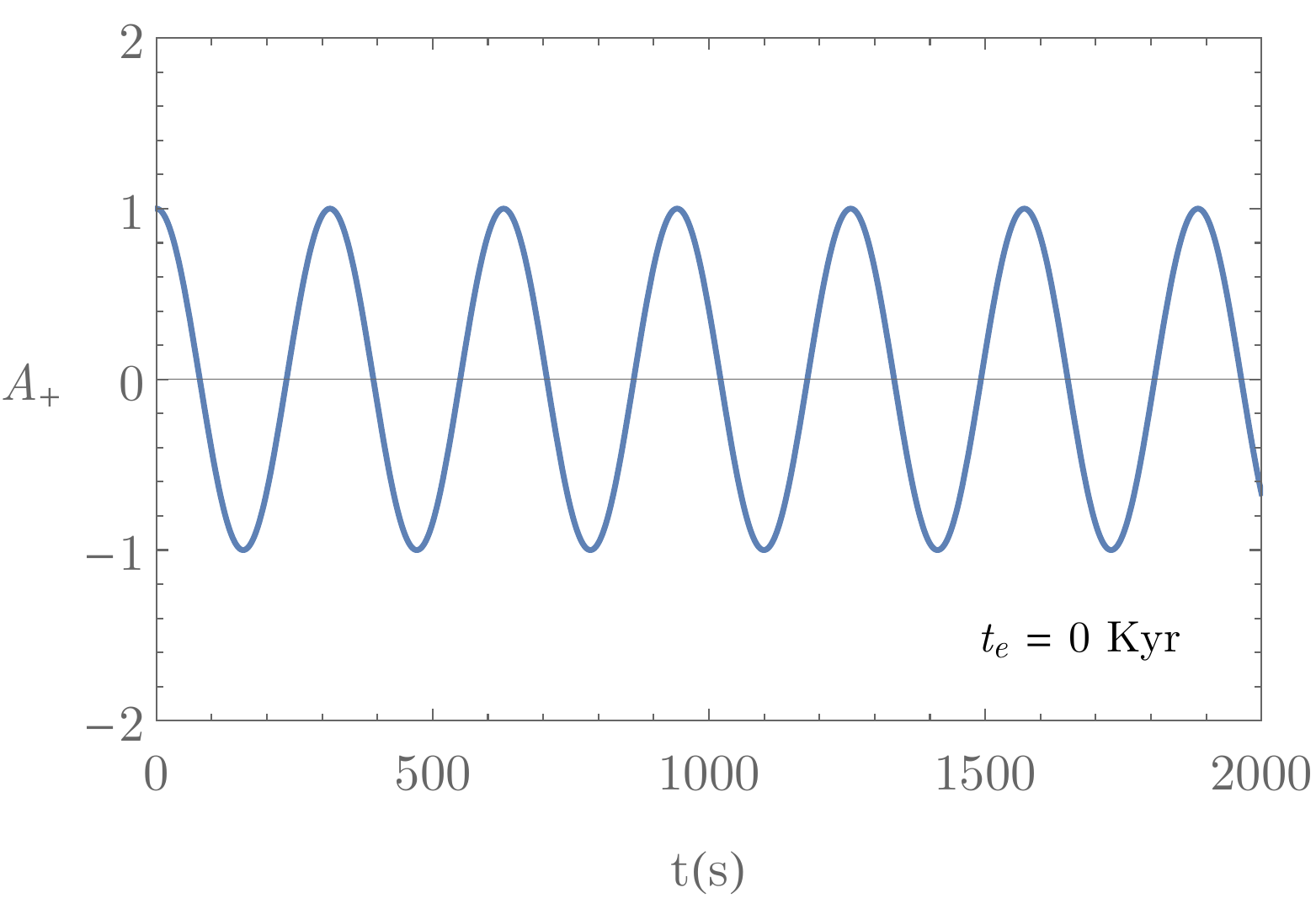}
\end{figure}
\begin{figure}[h!]
\includegraphics[scale=0.5]{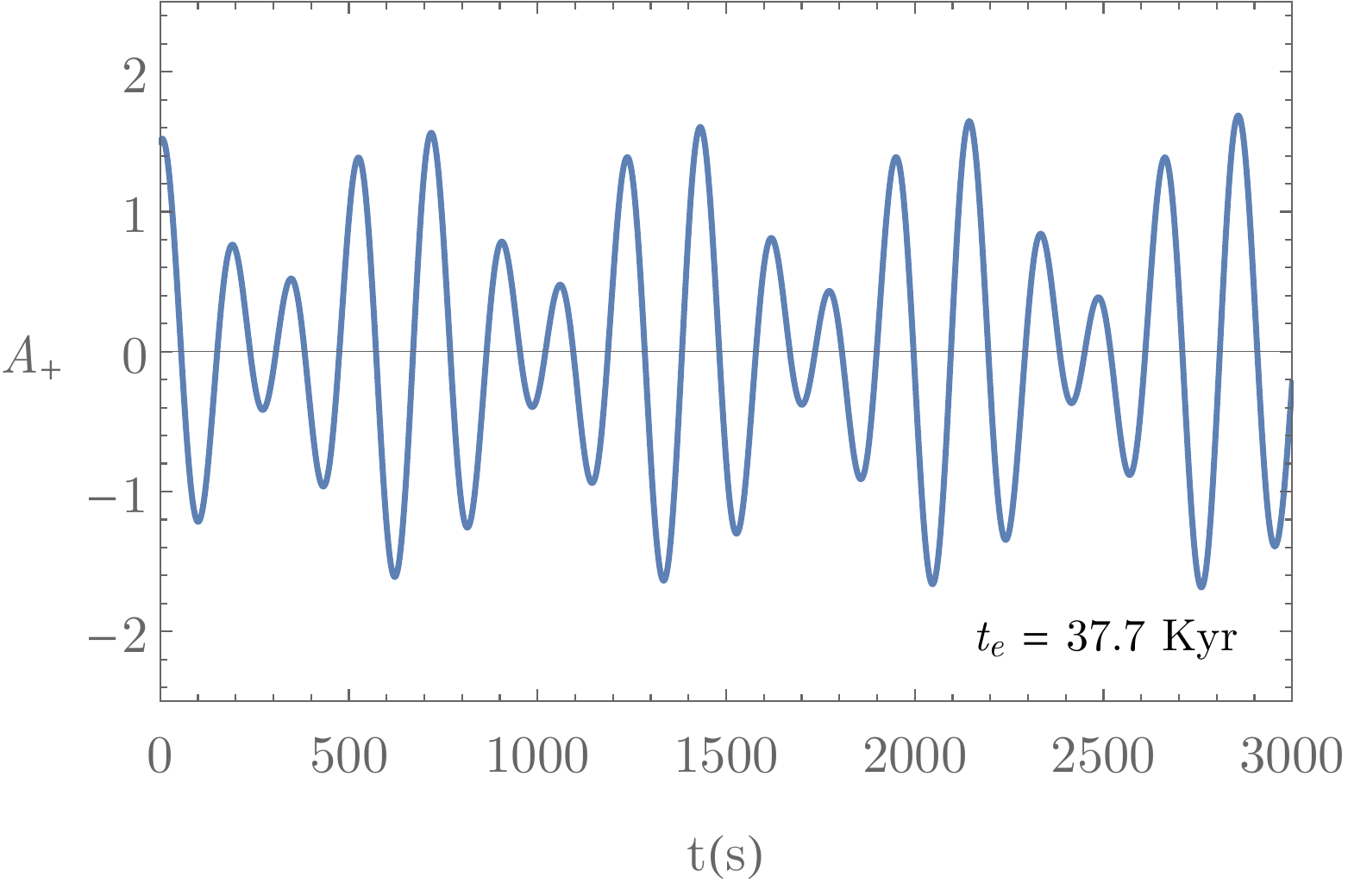}
\end{figure}
\begin{figure}[h!]
\includegraphics[scale=0.5]{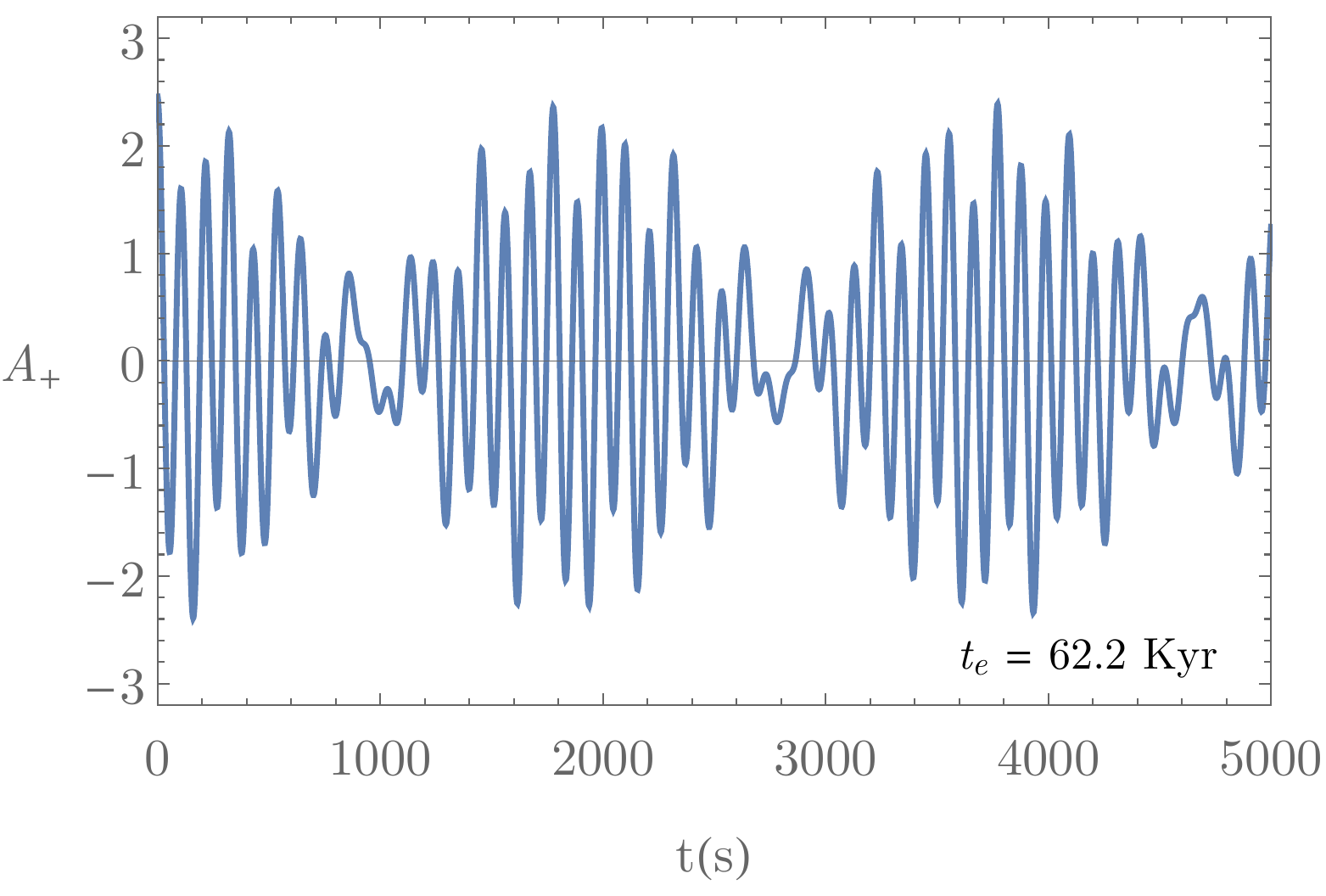}
\caption{Waveforms $\Theta_{+}=\tilde{h}_{+}+\Psi_{+}^{\left( 1\right)
}+\Psi_{+}^{\left( 2\right) }$ normalized by the amplitude $%
A_{+}=2r^{-1}c^{-4}\left( GM_{c}\right) ^{\frac{5}{3}}\left( 1+\cos ^{2}%
\protect\theta\right) \left( \protect\omega_{s}^{e}\right) ^{\frac{2}{3}}$
associated with the $\tilde{h}_{+}$ mode. The first plot presents only the
massless mode with frequency $\protect\omega_{s}^{e}=10^{-2}$ $Hz$. In the
second plot, we have a massless frequency mode $\protect\omega%
_{s}^{e}=1.762\times10^{-2}$ $Hz$ and two massive frequency modes $\protect%
\omega_{s}^{e_{m\left( 1\right) }}=1.326\times 10^{-2}$ $Hz$ and $\protect%
\omega_{s}^{e_{m\left( 2\right) }}=1.323\times10^{-2}$ $Hz$. In the third
plot, the massless mode has frequency $\protect\omega_{s}^{e}=3\times10^{-2}$
$Hz$ and the two massive modes have frequencies $\protect\omega%
_{s}^{e_{m\left( 1\right) }}=2.83\times10^{-2}$ $Hz$ and $\protect\omega %
_{s}^{e_{m\left( 2\right) }}=1.07\times10^{-2}$ $Hz$. All frequencies were
calculated from the solution of the balance equation (\protect\ref{BalanceEq}%
).}
\label{fig4}
\end{figure}

The plots in Fig. \ref{fig4} show the full tensorial mode waveforms
associated with polarization plus. In the interval $0\leq t_{e}<37.7$ $Kyr$
only the massless mode is present, and $\Theta_{+}$ has the form of a pure
sinusoid. From $t_{e}\geq$ $37.7$ $Kyr$, the three modes $\tilde{h}_{+}$, $%
\Psi_{+}^{\left( 1\right) }$ and $\Psi_{+}^{\left( 2\right) }$ combine and
generate an interference pattern. For $t_{e}\approx37.7$ $Kyr$ (second plot
of figure $4$), the two massive modes have similar frequencies and
contribute similarly to the structure of $\Theta_{+}$. As $t_{e}$ increases,
the $\Psi_{+}^{\left( 2\right) }$ mode (lower frequency mode) decreases in
importance compared to $\Psi_{+}^{\left( 1\right) }$. In the last plot of
figure \ref{fig4}, we present a moment when the complete waveform is mainly
characterized by $\tilde{h}_{+}$ and $\Psi_{+}^{\left( 1\right) }$, both
with similar frequencies. In this case, we notice that $\Theta_{+}$ is
modulated by an angular frequency envelope $\omega_{s}^{e}-\omega_{s}^{e_{m%
\left( 1\right) }}$ with some deformity generated by the presence of $%
\Psi_{+}^{\left( 2\right) }$.

In the next section, we will use the nonexistence of an interference
pattern in gravitational wave observations to constrain the parameter $%
m_{\Psi}$.

\subsection{Observational constraints\label{sec const}}

The waveform detected in the inspiral phase of a binary black hole system
allows constraining the parameter $m_{\Psi}$. It is possible because the
presence of massive modes produces an interference pattern (plots in Fig. $%
4$) that is clearly not observed. To obtain these constraints, we will
consider the event GW170104, which consists of the merge of two black holes
with a Chirp mass $M_{c}=21.1$ $M_{\odot}$ and a distance of $r=880$ $Mpc$
\cite{PhysRevLett.118.221101}.\footnote{%
We only take into account the best-fit parameters.} Figure \ref{fig1} of
Ref. \cite{PhysRevLett.118.221101} indicates that the gravitational wave is
detected initially at $t_{ini}=0.5$ $s$ with frequency $f_{GW}\approx45$ $Hz$%
, and the merge occurs around $t_{mer}=0.59$ $s$.

The estimated constraint for $m_{\Psi}$ is performed as follows:

\begin{itemize}
\item Using the initial condition $\omega_{s}\left( t_{ini}\right) =$ $\pi
f_{GW}=141$ $H_{z}$, we numerically solve the balance equation (\ref%
{BalanceEq}) for different values of $m_{\Psi}$. It is done considering that
the spin-$0$ mode is in the damping regime.

\item Knowing $\omega_{s}\left( t\right) $ and using $t_{e}=t_{ini}$, we
determine the largest value of $m_{\Psi}$ in which Eq. (\ref{Eq
algebrica tem}) has a solution for $t_{e_{m}}$. Numerically this corresponds
to a configuration equivalent to the second plot in Fig. \ref{fig3}.
\end{itemize}
The procedure above establishes the maximum value of $m_{\Psi }$ at which
the massive mode produces an interference pattern of the type shown in
Fig. \ref{fig4}. As this pattern is not observed in event GW170104, the
maximum value obtained establishes a lower bound for $m_{\Psi }$. Table \ref%
{table1} shows the result of this constraint:

\begin{table}[h!]
\begin{centering}
\begin{tabular}
[c]{|c|c|c|c|c|}\hline
& $\omega_{s}^{e}$ & $\omega_{s}^{e_{m}}$ & $m_{\Psi}\left(  \gtrsim\right)  $
& $\alpha\left(  \lesssim\right)  $\\\hline
Spin-$2$ & $141$ $Hz$ & $77$ $Hz$ & $4.2\times10^{-11}\text{ }m^{-1}$ &
$1.1\times10^{21}\text{ }m^{2}$\\\hline
\end{tabular}
\par\end{centering}
\caption{Emission frequencies of the massive ($\protect\omega _{s}^{e_{m}}$)
and massless ($\protect\omega _{s}^{e}$) tensorial modes, and constraints
for the parameters $m_{\Psi }$ and $\protect\alpha $ where $m_{\Psi }^{2}=2/%
\protect\alpha $. Note that the two massive modes have approximately the
same frequency given by $\protect\omega _{s}^{e_{m}}$. Furthermore, since $%
\protect\omega _{s}^{e}$ and $\protect\omega _{s}^{e_{m}}$ differ only by a
factor of $2$, the amplitudes of the modes $\Psi _{+.\times }^{\left(
1,2\right) }$ and $\tilde{h}_{+.\times }$ are similar. Thus, for $m_{\Psi
}<4.2\times 10^{-11}$ $m^{-1}$, we obtain an interference pattern similar to
the one shown in the second plot of figure \ref{fig4}.}
\label{table1}
\end{table}

Another way to obtain a more restrictive constraint for $m_{\Psi}$ is
through the coalescence time $ \Delta_{col}$. From Fig. \ref {fig1},
we see that the coalescence time increases substantially as $m_{\Psi }c$
decreases. On the other hand, event GW170104 observes a coalescence time $%
\Delta _{{col}}^{ob}\sim 0.1$ $s$ counted from the initial setting $%
\omega _{s}\left( t_{ini}\right) =141$ $H_{z}$. Thus, imposing that the
presence of the massive mode cannot change (in order of magnitude) the
observed value of $\Delta _{{col}}^{ob}$, we can constrain the
parameters $m_{\Psi }$ and $\alpha $.

The coalescence time is calculated from Eq. (\ref{BalanceEq}) and
essentially depends on the parameters $M_{c}$ and $m_{\Psi}$.\footnote{%
Again, we consider that the spin-$0$ mode is in the damping regime.}
Solutions to this equation show that as $M_{c}$ and $m_{\Psi}$ increase $%
\Delta_{{col}}$ decreases. In principle, any decrease in $m_{\Psi}$ can
be offset by an increase in $M_{c}$ leaving $\Delta_{{col}}$ invariant.
However, there is strong theoretical \cite{Woosley_2017} and observational
\cite{Fishbach_2017,obs} evidence that the mass of Chirp is "always" less
than $100M_{\odot}$.\footnote{%
By "always" we mean that the existence of a binary black hole system with $%
M_{c}\gtrsim100M_{\odot}$ is highly improbable.} Thus, assuming the maximum
value $M_{c}=100M_{\odot}$, we can establish a lower bound for $m_{\Psi}$
requiring $\Delta_{{col}}$ to be compatible with the observed value $%
\Delta_{{col}}^{ob}$.

In Fig. \ref{fig5}, we show the solution $\omega_{s}\left( t\right) $ for
different values of the parameter $m_{\Psi}$ with $M_{c}=100M_{\odot}$:
\begin{figure}[h!]
\includegraphics[scale=0.45]{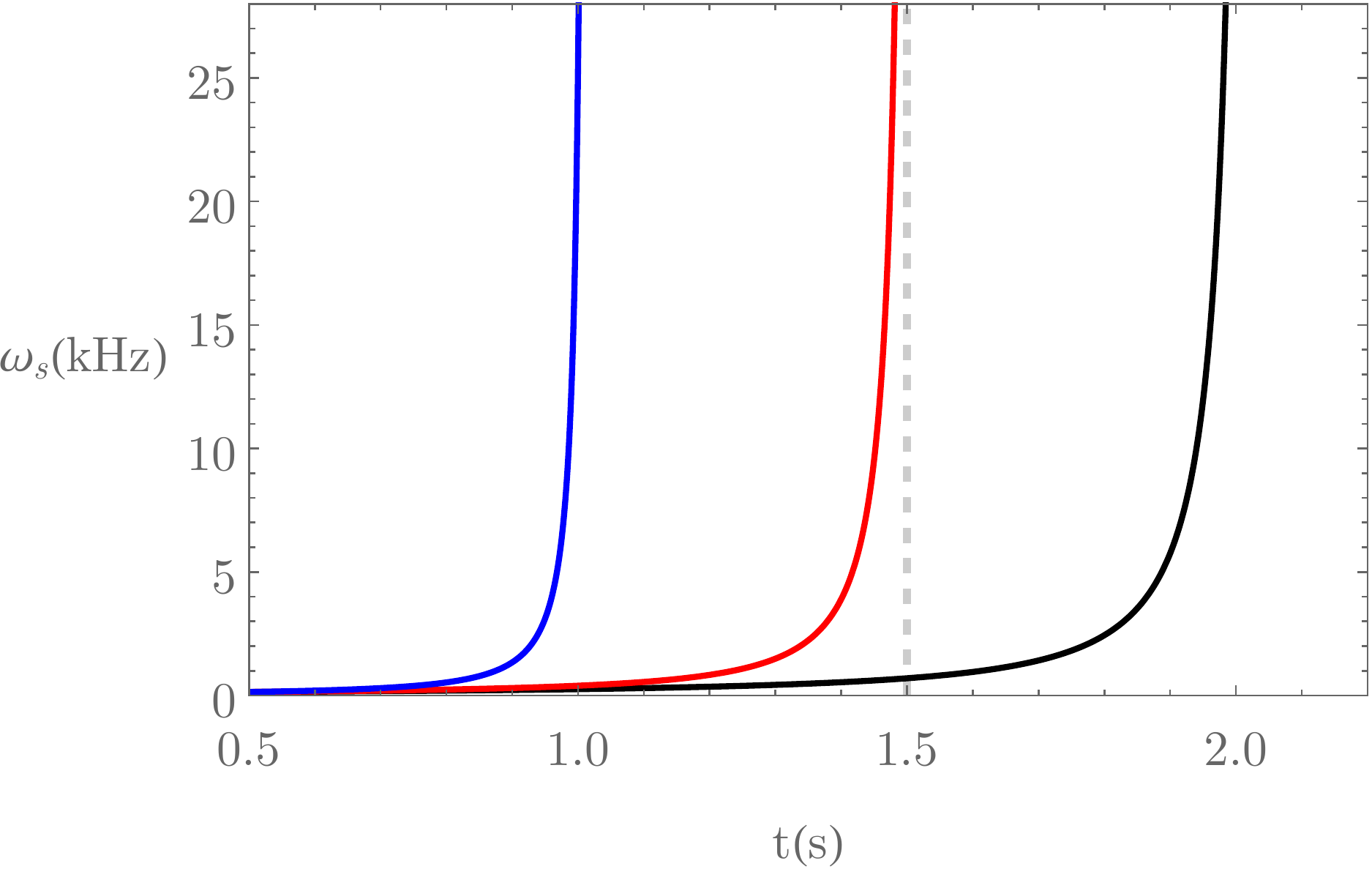}
\caption{Plot of the function $\protect\omega_{s}\left( t\right) $
considering the initial condition $\protect\omega_{s}\left( 0.5\right) =141$
$Hz$ and $M_{c}=100M_{\odot}$. The blue, red and black curves are
constructed with $m_{\Psi}c=120$ $s^{-1}$, $m_{\Psi}c=86$ $s^{-1}$ and $%
m_{\Psi}c=70$ $s^{-1}$, respectively. The vertical dashed curve indicates $%
\Delta_{{col}}=1$ $s$.}
\label{fig5}
\end{figure}

The coalescence time is calculated from $\Delta_{{col}}=t_{{col}%
}-0.5$, where $t_{{col}}$ is the instant when the functions in Fig. \ref{fig5} diverge. Thus, for the models described by the blue, red and
black curves, we obtain $\Delta_{{col}}=0.5$ $s$, $1$ $s$ and $1.5$ $s$%
, respectively. Furthermore, these values correspond to the minimum
coalescence times of a given model $m_{\Psi}$ as they were calculated with
the maximum value $M_{c}=100M_{\odot}$. If we decrease the chirp mass with $%
m_{\Psi}$ fixed, we get a larger value for $\Delta _{{col}}$.

Based on the previous discussion, we can estimate a lower bound for the
parameter $m_{\Psi}$ considering $\Delta_{{col}}\leq$ $\Delta_{{col%
}}^{ob}$. In principle, we should use $\Delta _{{col}}^{ob}\sim0.1$ $s$%
. However, to compensate for the errors generated by the approximations
performed (nonrelativistic dynamics, point masses, etc.), we reduce the
constraint by $1$ order of magnitude and consider $\Delta_{{col}}\leq1$
$s$.\footnote{%
Relativistic corrections are essential to correctly describe the orbital
dynamics close to the merger. However, they do not change the coalescence
time in order of magnitude. For example, the predicted coalescence time for
event GW1701104 in GR on the quadrupole approximation is $\Delta_{{col}%
}=0.16$ $s$.} Thus, from Fig. \ref{fig5}, we see that $m_{\Psi}c$ must be
larger than $86$ $s^{-1}$, which results in the following constraint shown
in Table \ref{table2}.
\begin{table}[h!]
\begin{centering}
\begin{tabular}
[c]{|c|c|c|}\hline
& $m_{\Psi}\left(  \gtrsim\right)  $ & $\alpha\left(  \lesssim\right)
$\\\hline
Spin-$2$ & $3\times10^{-7}\text{ }m^{-1}$ & $2\times10^{13}\text{ }m^{2}%
$\\\hline
\end{tabular}
\par\end{centering}
\caption{Constraints to parameters $m_{\Psi}$ and $\protect\alpha$ taking
into account $\Delta_{{col}}\leq1$ $s$.}
\label{table2}
\end{table}

\bigskip Due to the various approximations performed, the estimates
presented in Table \ref{table2} must be considered only in order of
magnitude. Still, the constraint for the $\alpha $ parameter via coalescence
time is about $8$ orders of magnitude more restrictive than the constraint
obtained via the interference pattern.

\section{Final Comments\label{sec final}}

In this paper, we studied GWs emitted by binary point-mass black hole
systems in the context of the quadratic gravity model assuming
nonrelativistic and \ circular orbit approximations. The GWs solutions were
calculated by taking into account the dominant terms in the multipolar
expansion. They exhibit three modes: a massive spin-$2$ mode $\Psi_{\mu\nu}$%
, a massive spin-$0$ mode $\Phi$, and the expected massless spin-$2$ mode $%
\tilde{h}_{\mu\nu}$. Besides, the massive modes present two different
behaviors: the oscillatory regime, which plays the real role of GW; and the
damped regime, which presents an exponential decay.

We calculated the energy-momentum tensor $t_{\mu\nu}$ of the GWs and showed
that it presents the Ostrogradsky instability. To circumvent this potential
problem and obtain a consistent physical interpretation \cite{Edelstein_2021}, we consider spin-$2$
waves as a single structure given as a result of the destructive
interferences between the fields $\tilde{h}_{\mu\nu}$ and $\Psi_{\mu\nu}$.
Having obtained the $t_{\mu\nu}$, we constructed the energy balance equation
and studied how the quadratic gravity model modifies the orbital dynamics of
point-mass black hole binaries. We showed that the spin-$2$ structure takes
a longer time to reach coalescence indicating the system loses energy more
slowly when compared to GR. For the case of spin-$0$ plus GR, we have a
shorter time to reach coalescence indicating that the system loses energy
faster than the GR pure case.

We determined the waveform that a detector would observe for the complete
spin-$2$ structure. When the massive mode is not present, we obtain a pure
sine wave. However, when it is present, we get a clear interference pattern.
Using the nonobservation of this interference pattern, we constrain the
parameter to $m_{\Psi}\gtrsim4.2\times10^{-11}m^{-1}$ or $\alpha
\lesssim1.1\times10^{21}m^{2}$. Furthermore, based on the coalescence time,
we developed a second method to constrain $m_{\Psi}$. By this method, we
obtain $m_{\Psi}\gtrsim3\times10^{-7}m^{-1}$ or $\alpha\lesssim2%
\times10^{13}m^{2}$, which is $8$ orders of magnitude more restrictive than
the previous one.

The methods presented in Sec. \ref{sec const} do not effectively
constrain the parameter $m_{\Phi}$ or $\gamma$. The main reason for this is the
structural difference between the solutions $\Phi$ and $\Psi_{+,\times}$. For example, the scalar mode presents a longitudinal polarization differently from the tensor modes \cite{capozziello2006scalar,CORDA_2008,Capozziello_2008}.
This difference leads to factor $1/18$ which appears in the balance equation
and makes the energy loss via scalar mode\ ineffective.

There are some papers in the literature that constrain the parameters of
quadratic gravity. On astrophysical scales, of the order of stellar systems,
we get restrictions associated with the scalar mode arising from binary
systems. Using observations of decreasing orbital period of neutron star
binaries, Refs. \cite%
{PhysRevD.84.024027,PhysRevD.83.104022} and \cite{PhysRevD.104.084061} constrain $\gamma\lesssim10^{17}m^{2}$
and $\gamma\lesssim10^{16}m^{2}$, respectively. At earthly scales,
the E\"{o}t-Wash torsion balance experiment \cite{PhysRevLett.98.021101} provides bounds for both
$\gamma$ and $\alpha$ parameters. Based on the static weak-field solution of
quadratic gravity \cite{PhysRevD.16.953}, the E\"{o}t-Wash experiment constrains $\alpha
\sim\gamma\lesssim2\times10^{-9}m^{2}$. Even though earthly bounds are much
more restrictive than astrophysical ones, it is always important to test
alternative models of gravity at different scales.

The circular orbit approximation motivates future work in which noncircular
orbits are considered from the beginning. In this context, it would be
possible to study the loss of angular momentum and the effects of orbit
circularization. This would be particularly interesting in quadratic
gravity, where the presence of the Weyl-Weyl term generates the Ostrogradsky
instability.

\begin{acknowledgments}
L.G.M. is grateful to CNPq-Brazil (Grant No. 308380/2019-3) for partial
financial support. M.F.S.A. acknowledges CNPq-Brazil for financial support. L.F.M.A.M.R. thanks CAPES-Brazil for financial support.
\end{acknowledgments}

\appendix
\begin{widetext}
\section{The gravitational energy-momentum tensor\label{sec - Ap GraEneMoTen}%
}
In this Appendix, we calculate the gravitational energy-momentum tensor%
\begin{equation}
t_{\mu\alpha}=-\frac{c^{4}}{8\pi G}\left[  \left\langle G_{\mu\alpha}^{\left(
2\right)  }\right\rangle +\gamma\left\langle H_{\mu\alpha}^{\left(  2\right)
}\right\rangle -2\alpha\left\langle I_{\mu\alpha}^{\left(  2\right)
}\right\rangle \right]  \label{tmualpha}%
\end{equation}
where $G_{\mu\alpha}^{\left(  2\right)  }$, $H_{\mu\alpha}^{\left(  2\right)
}$ and $I_{\mu\alpha}^{\left(  2\right)  }$ are given in Eqs. (\ref{Gmunu2}),
(\ref{Hmunu2}), and (\ref{Imunu2}), respectively.

The brackets $\left\langle ...\right\rangle $ represent space-time averages of
wave-like solutions obtained from Eqs. (\ref{Eq field lin 1}),
(\ref{Eq field lin 2}), and (\ref{Eq field lin 3}). These averages are taken
considering several periods of time and space in such a way that in these
averages we can perform integrations by parts and neglect the surface terms
\cite{magiorrebook}. Using this property, we will simplify $\left\langle
H_{\mu\alpha}^{\left(  2\right)  }\right\rangle $ and $\left\langle
I_{\mu\alpha}^{\left(  2\right)  }\right\rangle $. So, remembering that%
\[
\nabla_{\mu}\partial_{\alpha}R=\partial_{\mu}\partial_{\alpha}R-\Gamma_{\text{
\ }\mu\alpha}^{\rho}\partial_{\rho}R,
\]
we have for $\left\langle H_{\mu\alpha}^{\left(  2\right)  }\right\rangle $
\begin{align*}
\left\langle H_{\mu\alpha}^{\left(  2\right)  }\right\rangle  &  =\left\langle
R^{\left(  1\right)  }R_{\mu\alpha}^{\left(  1\right)  }\right\rangle
-\frac{1}{4}\eta_{\mu\alpha}\left\langle R^{\left(  1\right)  }R^{\left(
1\right)  }\right\rangle +\left\langle h_{\mu\alpha}\square R^{\left(
1\right)  }\right\rangle +\eta_{\mu\alpha}\left\langle \left[  g^{\lambda
\kappa}\nabla_{\kappa}\partial_{\lambda}\right]  ^{\left(  1\right)
}R^{\left(  1\right)  }\right\rangle +\left\langle \Gamma_{\text{ \ }\mu
\alpha}^{\rho\left(  1\right)  }\partial_{\rho}R^{\left(  1\right)
}\right\rangle \\
&  =\left\langle R^{\left(  1\right)  }R_{\mu\alpha}^{\left(  1\right)
}\right\rangle -\frac{1}{4}\eta_{\mu\alpha}\left\langle R^{\left(  1\right)
}R^{\left(  1\right)  }\right\rangle +\left\langle h_{\mu\alpha}\square
R^{\left(  1\right)  }\right\rangle -\eta_{\mu\alpha}\left\langle
h^{\lambda\kappa}\partial_{\kappa}\partial_{\lambda}R^{\left(  1\right)
}\right\rangle -\eta_{\mu\alpha}\eta^{\lambda\kappa}\left\langle
\Gamma_{\text{ \ }\kappa\lambda}^{\rho\left(  1\right)  }\partial_{\rho
}R^{\left(  1\right)  }\right\rangle \\
&  +\left\langle \Gamma_{\text{ \ }\mu\alpha}^{\rho\left(  1\right)  }%
\partial_{\rho}R^{\left(  1\right)  }\right\rangle ,
\end{align*}
where $\square=$ $\partial^{\sigma}\partial_{\sigma}$.To obtain $\left\langle
I_{\mu\alpha}^{\left(  2\right)  }\right\rangle $, we must develop the term%
\begin{align*}
\left\langle \left(  \nabla^{\nu}\nabla^{\beta}\right)  ^{\left(  1\right)
}C_{\mu\nu\alpha\beta}^{\left(  1\right)  }\right\rangle  &  =\left\langle
\left(  g^{\nu\kappa}\partial_{\kappa}\nabla^{\beta}\right)  ^{\left(
1\right)  }C_{\mu\nu\alpha\beta}^{\left(  1\right)  }\right\rangle
-\left\langle \left(  g^{\nu\kappa}\Gamma_{\text{ \ }\kappa\mu}^{\rho}%
\nabla^{\beta}\right)  ^{\left(  1\right)  }C_{\rho\nu\alpha\beta}^{\left(
1\right)  }\right\rangle -\left\langle \left(  g^{\nu\kappa}\Gamma_{\text{
\ }\kappa\nu}^{\rho}\nabla^{\beta}\right)  ^{\left(  1\right)  }C_{\mu
\rho\alpha\beta}^{\left(  1\right)  }\right\rangle \\
&  -\left\langle \left(  g^{\nu\kappa}\Gamma_{\text{ \ }\kappa\alpha}^{\rho
}\nabla^{\beta}\right)  ^{\left(  1\right)  }C_{\mu\nu\rho\beta}^{\left(
1\right)  }\right\rangle \\
&  =-\eta^{\beta\lambda}\left\langle h^{\nu\kappa}\partial_{\kappa}%
\partial_{\lambda}C_{\mu\nu\alpha\beta}^{\left(  1\right)  }\right\rangle
-\eta^{\nu\kappa}\left\langle \Gamma_{\text{ \ }\kappa\mu}^{\rho\left(
1\right)  }\partial^{\beta}C_{\rho\nu\alpha\beta}^{\left(  1\right)
}\right\rangle -\eta^{\nu\kappa}\left\langle \Gamma_{\text{ \ }\kappa\nu
}^{\rho\left(  1\right)  }\partial^{\beta}C_{\mu\rho\alpha\beta}^{\left(
1\right)  }\right\rangle \\
&  -\eta^{\nu\kappa}\left\langle \Gamma_{\text{ \ }\kappa\alpha}^{\rho\left(
1\right)  }\partial^{\beta}C_{\mu\nu\rho\beta}^{\left(  1\right)
}\right\rangle .
\end{align*}
So, $I_{\mu\alpha}^{\left(  2\right)  }$ is written as%
\begin{align*}
\left\langle I_{\mu\alpha}^{\left(  2\right)  }\right\rangle  & =-\left\langle
h^{\nu\kappa}\partial_{\kappa}\partial^{\beta}C_{\mu\nu\alpha\beta}^{\left(
1\right)  }\right\rangle -\eta^{\nu\kappa}\left\langle \Gamma_{\text{
\ }\kappa\mu}^{\rho\left(  1\right)  }\partial^{\beta}C_{\rho\nu\alpha\beta
}^{\left(  1\right)  }\right\rangle -\eta^{\nu\kappa}\left\langle
\Gamma_{\text{ \ }\kappa\nu}^{\rho\left(  1\right)  }\partial^{\beta}%
C_{\mu\rho\alpha\beta}^{\left(  1\right)  }\right\rangle -\eta^{\nu\kappa
}\left\langle \Gamma_{\text{ \ }\kappa\alpha}^{\rho\left(  1\right)  }%
\partial^{\beta}C_{\mu\nu\rho\beta}^{\left(  1\right)  }\right\rangle \\
& +\frac{1}{2}\eta^{\nu\rho}\eta^{\beta\lambda}\left\langle R_{\rho\lambda
}^{\left(  1\right)  }C_{\mu\nu\alpha\beta}^{\left(  1\right)  }\right\rangle
.
\end{align*}
\ Therefore,
\begin{align}
\left\langle G_{\mu\alpha}^{\left(  2\right)  }\right\rangle  &  =\left\langle
R_{\mu\alpha}^{\left(  2\right)  }\right\rangle -\frac{1}{2}\eta_{\mu\alpha
}\left\langle R^{\left(  2\right)  }\right\rangle -\frac{1}{2}\left\langle
h_{\mu\alpha}R^{\left(  1\right)  }\right\rangle ,\label{G_munu 2}\\
\left\langle H_{\mu\alpha}^{\left(  2\right)  }\right\rangle  &  =\left\langle
R^{\left(  1\right)  }R_{\mu\alpha}^{\left(  1\right)  }\right\rangle
-\frac{1}{4}\eta_{\mu\alpha}\left\langle R^{\left(  1\right)  }R^{\left(
1\right)  }\right\rangle +\left\langle h_{\mu\alpha}\square R^{\left(
1\right)  }\right\rangle -\eta_{\mu\alpha}\left\langle h^{\lambda\kappa
}\partial_{\kappa}\partial_{\lambda}R^{\left(  1\right)  }\right\rangle
-\eta_{\mu\alpha}\eta^{\lambda\kappa}\left\langle \Gamma_{\text{ \ }%
\kappa\lambda}^{\rho\left(  1\right)  }\partial_{\rho}R^{\left(  1\right)
}\right\rangle \label{H_munu 2}\\
&  +\left\langle \Gamma_{\text{ \ }\mu\alpha}^{\rho\left(  1\right)  }%
\partial_{\rho}R^{\left(  1\right)  }\right\rangle ,\nonumber\\
\left\langle I_{\mu\alpha}^{\left(  2\right)  }\right\rangle  &
=-\left\langle h^{\nu\kappa}\partial_{\kappa}\partial^{\beta}C_{\mu\nu
\alpha\beta}^{\left(  1\right)  }\right\rangle -\eta^{\nu\kappa}\left\langle
\Gamma_{\text{ \ }\kappa\mu}^{\rho\left(  1\right)  }\partial^{\beta}%
C_{\rho\nu\alpha\beta}^{\left(  1\right)  }\right\rangle -\eta^{\nu\kappa
}\left\langle \Gamma_{\text{ \ }\kappa\nu}^{\rho\left(  1\right)  }%
\partial^{\beta}C_{\mu\rho\alpha\beta}^{\left(  1\right)  }\right\rangle
-\eta^{\nu\kappa}\left\langle \Gamma_{\text{ \ }\kappa\alpha}^{\rho\left(
1\right)  }\partial^{\beta}C_{\mu\nu\rho\beta}^{\left(  1\right)
}\right\rangle \label{I_munu 2}\\
&  +\frac{1}{2}\eta^{\nu\rho}\eta^{\beta\lambda}\left\langle R_{\rho\lambda
}^{\left(  1\right)  }C_{\mu\nu\alpha\beta}^{\left(  1\right)  }\right\rangle
.\nonumber
\end{align}

The next step is to calculate the various terms needed using decomposition
(\ref{Decomposicao metrica}). As $t_{\mu\nu}$ is calculated in vacuum, we can
use the traceless-transverse gauge, and in this case, $\tilde{h}=\Psi=0$ e
$\square\tilde{h}_{\mu\alpha}=0$. Then, the decomposed metric is rewritten as%
\begin{equation}
\bar{h}_{\mu\alpha}=\tilde{h}_{\mu\alpha}+\Psi_{\mu\alpha}-\eta_{\mu\alpha
}\Phi. \label{h decomposto}%
\end{equation}
With these simplifications, the first-order quantities obtained in Sec.
\ref{sec - weak field} simplify to%

\begin{equation}
R^{\left(  1\right)  }\approx-3\square\Phi, \label{R 1 decomposto}%
\end{equation}%
\begin{equation}
R_{\mu\alpha}^{\left(  1\right)  }\approx-\frac{1}{2}\square\Psi_{\mu\alpha
}-\partial_{\mu}\partial_{\alpha}\Phi+\eta_{\mu\alpha}\square\Phi-\eta
_{\mu\alpha}\frac{\Phi}{2\gamma}, \label{Ricci 1 decomposta}%
\end{equation}%
\begin{align}
C_{\mu\nu\alpha\beta}^{\left(  1\right)  }  &  =\frac{1}{2}\left[
\partial_{\alpha}\partial_{\nu}\left(  \tilde{h}_{\mu\beta}+\Psi_{\mu\beta
}\right)  -\partial_{\beta}\partial_{\nu}\left(  \tilde{h}_{\mu\alpha}%
+\Psi_{\mu\alpha}\right)  +\partial_{\beta}\partial_{\mu}\left(  \tilde
{h}_{\nu\alpha}+\Psi_{\nu\alpha}\right)  -\partial_{\alpha}\partial_{\mu
}\left(  \tilde{h}_{\nu\beta}+\Psi_{\nu\beta}\right)  \right] \nonumber\\
&  +\frac{1}{4}\left[  \eta_{\mu\alpha}\square\Psi_{\nu\beta}-\eta_{\mu\beta
}\square\Psi_{\nu\alpha}+\eta_{\nu\beta}\square\Psi_{\mu\alpha}-\eta
_{\nu\alpha}\square\Psi_{\mu\beta}\right]  , \label{Weyl 1 decomposto}%
\end{align}%
\begin{equation}
\partial^{\beta}C_{\mu\nu\alpha\beta}^{\left(  1\right)  }=\frac{1}{4}\left[
\partial_{\mu}\square\Psi_{\nu\alpha}-\partial_{\nu}\square\Psi_{\mu\alpha
}\right]  , \label{Der Weyl 1 decomposto}%
\end{equation}
and%
\begin{equation}
\Gamma_{\, \text{\ \ }\kappa\mu}^{\rho\left(  1\right)  }=\frac{1}{2}%
\eta^{\rho\lambda}\left(  \partial_{\kappa}h_{\lambda\mu}+\partial_{\mu
}h_{\lambda\kappa}-\partial_{\lambda}h_{\kappa\mu}\right)  .
\label{Conexao 1 decomposta}%
\end{equation}

The term $\left\langle G_{\mu\alpha}^{\left(  2\right)  }\right\rangle $ is
given by the Eq. (\ref{G_munu 2}). The calculation of the terms $\left\langle
R_{\mu\alpha}^{\left(  2\right)  }\right\rangle $ and $\left\langle R^{\left(
2\right)  }\right\rangle $ are extensive but straightforward and results in%
\begin{align}
\left\langle R_{\mu\alpha}^{\left(  2\right)  }\right\rangle  & =-\frac{1}%
{4}\left\langle \partial_{\mu}\tilde{h}_{\nu\beta}\partial_{\alpha}\tilde
{h}^{\nu\beta}\right\rangle -\frac{1}{4}\left\langle \partial_{\mu}\Psi
_{\nu\beta}\partial_{\alpha}\Psi^{\nu\beta}\right\rangle +\frac{1}%
{2}\left\langle \partial_{\alpha}\Phi\partial_{\mu}\Phi\right\rangle -\frac
{1}{2}\left\langle \partial_{\mu}\tilde{h}_{\nu\beta}\partial_{\alpha}%
\Psi^{\nu\beta}\right\rangle +\frac{1}{2}\left\langle \partial_{\lambda}%
\Psi_{\text{ \ }\mu}^{\nu}\partial^{\lambda}\Psi_{\nu\alpha}\right\rangle
\label{R_munu 2 media}\\
& -\frac{1}{2}\eta_{\alpha\mu}\left\langle \partial^{\lambda}\Phi
\partial_{\lambda}\Phi\right\rangle ,\nonumber
\end{align}
and%
\begin{equation}
\left\langle R^{\left(  2\right)  }\right\rangle =-\frac{9}{2}\left\langle
\partial^{\lambda}\Phi\partial_{\lambda}\Phi\right\rangle -\frac{1}%
{4}\left\langle \partial_{\lambda}\Psi^{\nu\alpha}\partial^{\lambda}\Psi
_{\nu\alpha}\right\rangle .\label{R 2 media}%
\end{equation}
Substituting the results (\ref{R_munu 2 media}), (\ref{R 2 media}),
(\ref{R 1 decomposto}), and (\ref{h decomposto}) in $\left\langle G_{\mu\alpha
}^{\left(  2\right)  }\right\rangle $, we get
\begin{align}
\left\langle G_{\mu\alpha}^{\left(  2\right)  }\right\rangle  &  =-\frac{1}%
{4}\left\langle \partial_{\mu}\tilde{h}_{\nu\beta}\partial_{\alpha}\tilde
{h}^{\nu\beta}\right\rangle -\frac{1}{4}\left\langle \partial_{\mu}\Psi
_{\nu\beta}\partial_{\alpha}\Psi^{\nu\beta}\right\rangle -\frac{1}%
{2}\left\langle \partial_{\mu}\tilde{h}_{\nu\beta}\partial_{\alpha}\Psi
^{\nu\beta}\right\rangle +\frac{1}{2}\left\langle \partial_{\alpha}%
\Phi\partial_{\mu}\Phi\right\rangle \nonumber\\
&  -\frac{1}{2}\left\langle \Psi_{\text{ \ }\mu}^{\nu}\square\Psi_{\nu\alpha
}\right\rangle -\frac{1}{4}\eta_{\mu\alpha}\left\langle \Phi\square
\Phi\right\rangle -\frac{1}{8}\eta_{\mu\alpha}\left\langle \Psi^{\nu\beta
}\square\Psi_{\nu\beta}\right\rangle +\frac{3}{2}\left\langle \Psi_{\mu\alpha
}\square\Phi\right\rangle .\label{G_munu media}%
\end{align}
The field equations (\ref{Eq field lin 2}) and (\ref{Eq field lin 3}) in
vacuum can be written as
\begin{equation}
\square\Psi_{\mu\alpha}=\frac{1}{\alpha}\Psi_{\mu\alpha}\text{ \ \ and
\ \ }\square\Phi=\frac{1}{3\gamma}\Phi.\label{Eq field vacuo}%
\end{equation}
Substituting these equations into $\left\langle G_{\mu\alpha}^{\left(
2\right)  }\right\rangle $, we get%
\begin{align}
\left\langle G_{\mu\alpha}^{\left(  2\right)  }\right\rangle  &  =-\frac{1}%
{4}\left\langle \partial_{\mu}\tilde{h}_{\nu\beta}\partial_{\alpha}\tilde
{h}^{\nu\beta}\right\rangle -\frac{1}{4}\left\langle \partial_{\mu}\Psi
_{\nu\beta}\partial_{\alpha}\Psi^{\nu\beta}\right\rangle -\frac{1}%
{2}\left\langle \partial_{\mu}\tilde{h}_{\nu\beta}\partial_{\alpha}\Psi
^{\nu\beta}\right\rangle +\frac{1}{2}\left\langle \partial_{\alpha}%
\Phi\partial_{\mu}\Phi\right\rangle \nonumber\\
&  -\frac{1}{2\alpha}\left\langle \Psi_{\text{ \ }\mu}^{\nu}\Psi_{\nu\alpha
}\right\rangle -\frac{1}{12\gamma}\eta_{\mu\alpha}\left\langle \Phi
^{2}\right\rangle -\frac{1}{8\alpha}\eta_{\mu\alpha}\left\langle \Psi
^{\nu\beta}\Psi_{\nu\beta}\right\rangle ,\label{G_munu media 1}%
\end{align}
where we use that $\left\langle \Psi_{\mu\alpha}\Phi\right\rangle =0$. The
relation $\left\langle \Psi_{\mu\alpha}\Phi\right\rangle =0$ is derived from
the vacuum field equations from the following construction:%
\begin{gather*}
\Phi \square \Psi _{\mu \alpha }=\frac{1}{\alpha }\Phi \Psi _{\mu \alpha }%
\text{ \ and \ }\Psi _{\mu \alpha }\square \Phi =\frac{1}{3\gamma }\Phi \Psi
_{\mu \alpha }\Rightarrow  \\
\left\langle \Phi \square \Psi _{\mu \alpha }-\Psi _{\mu \alpha }\square
\Phi \right\rangle =\left( \frac{1}{\alpha }-\frac{1}{3\gamma }\right)
\left\langle \Phi \Psi _{\mu \alpha }\right\rangle \Rightarrow  \\
\frac{1}{3\gamma \alpha }\left( 3\gamma -\alpha \right) \left\langle \Phi
\Psi _{\mu \alpha }\right\rangle =0.
\end{gather*}
Thus, for $\alpha\neq3\gamma$, we have $\left\langle \Phi\Psi_{\mu\alpha
}\right\rangle =0$.
The quantity $\left\langle H_{\mu\alpha}^{\left(  2\right)  }\right\rangle $
is given by the Eq. (\ref{H_munu 2}). Calculating each of the terms present in
$\left\langle H_{\mu\alpha}^{\left(  2\right)  }\right\rangle $, we obtain

\begin{itemize}
\item First term:%
\begin{equation}
\left\langle R^{\left(  1\right)  }R_{\mu\alpha}^{\left(  1\right)
}\right\rangle =\frac{3}{2}\left\langle \square\Phi\square\Psi_{\mu\alpha
}\right\rangle +3\left\langle \square\Phi\partial_{\mu}\partial_{\alpha}%
\Phi\right\rangle -3\eta_{\mu\alpha}\left\langle \left(  \square\Phi\right)
^{2}\right\rangle +\eta_{\mu\alpha}\frac{3}{2\gamma}\left\langle \Phi
\square\Phi\right\rangle . \label{1 terrmo H}%
\end{equation}

\item Second term:%
\begin{equation}
\left\langle R^{\left(  1\right)  }R^{\left(  1\right)  }\right\rangle
=9\left\langle \left(  \square\Phi\right)  ^{2}\right\rangle .
\label{2 termo H}%
\end{equation}

\item Third term:%
\begin{equation}
\left\langle h_{\mu\alpha}\square R^{\left(  1\right)  }\right\rangle
=-3\left\langle \square\Psi_{\alpha\mu}\square\Phi\right\rangle -3\eta
_{\alpha\mu}\left\langle \left(  \square\Phi\right)  ^{2}\right\rangle .
\label{3 terrmo H}%
\end{equation}

\item Fourth term:%
\begin{equation}
\left\langle h^{\lambda\kappa}\partial_{\kappa}\partial_{\lambda}R^{\left(
1\right)  }\right\rangle =-3\left\langle \left(  \square\Phi\right)
^{2}\right\rangle . \label{4 termo H}%
\end{equation}

\item Fifth term:%
\begin{equation}
\eta^{\lambda\kappa}\left\langle \Gamma_{\text{ \ }\kappa\lambda}^{\rho\left(
1\right)  }\partial_{\rho}R^{\left(  1\right)  }\right\rangle =-3\left\langle
\left(  \square\Phi\right)  ^{2}\right\rangle . \label{5 terrmo H}%
\end{equation}

\item Sixth term:%
\begin{equation}
\left\langle \Gamma_{\text{ \ }\mu\alpha}^{\rho\left(  1\right)  }%
\partial_{\rho}R^{\left(  1\right)  }\right\rangle =3\left\langle
\partial_{\mu}\partial_{\alpha}\Phi\square\Phi\right\rangle -\frac{3}%
{2}\left\langle \square\Psi_{\alpha\mu}\square\Phi\right\rangle -\frac{3}%
{2}\eta_{\alpha\mu}\left\langle \left(  \square\Phi\right)  ^{2}\right\rangle
. \label{6 termo H}%
\end{equation}

\end{itemize}

Then we substitute these six results in the expression of $\left\langle
H_{\mu\alpha}^{\left(  2\right)  }\right\rangle $ obtaining%
\begin{equation}
\left\langle H_{\mu\alpha}^{\left(  2\right)  }\right\rangle =6\left\langle
\square\Phi\partial_{\mu}\partial_{\alpha}\Phi\right\rangle -\frac{15}{4}%
\eta_{\mu\alpha}\left\langle \left(  \square\Phi\right)  ^{2}\right\rangle
+\eta_{\mu\alpha}\frac{3}{2\gamma}\left\langle \Phi\square\Phi\right\rangle
-3\left\langle \square\Psi_{\alpha\mu}\square\Phi\right\rangle .
\label{H_munu media}%
\end{equation}
Analogously to the case of $\left\langle G_{\mu\alpha}^{\left(  2\right)
}\right\rangle $, we can use the field equations (\ref{Eq field vacuo}) and
rewrite $\left\langle H_{\mu\alpha}^{\left(  2\right)  }\right\rangle $ as%
\begin{equation}
\left\langle H_{\mu\alpha}^{\left(  2\right)  }\right\rangle =\frac{1}{\gamma
}\left(  \frac{1}{12\gamma}\eta_{\mu\alpha}\left\langle \Phi^{2}\right\rangle
-2\left\langle \partial_{\mu}\Phi\partial_{\alpha}\Phi\right\rangle \right)  ,
\label{H_munu media 1}%
\end{equation}
where we use again that $\left\langle \Psi_{\alpha\mu}\Phi\right\rangle =0$.

The quantity $\left\langle I_{\mu\alpha}^{\left(  2\right)  }\right\rangle $
is given by the Eq. (\ref{I_munu 2}). Calculating each of the terms present in
$\left\langle I_{\mu\alpha}^{\left(  2\right)  }\right\rangle $, we obtain:

\begin{itemize}
\item First term:%
\begin{equation}
\left\langle h^{\nu\kappa}\partial_{\kappa}\partial^{\beta}C_{\mu\nu
\alpha\beta}^{\left(  1\right)  }\right\rangle =-\frac{1}{4}\left\langle
\square\Phi\square\Psi_{\mu\alpha}\right\rangle . \label{1 termo I}%
\end{equation}

\item Second term:%
\begin{equation}
\eta^{\nu\kappa}\left\langle \Gamma_{\text{ \ }\kappa\mu}^{\rho\left(
1\right)  }\partial^{\beta}C_{\rho\nu\alpha\beta}^{\left(  1\right)
}\right\rangle =\frac{1}{4}\left\langle \square\Psi_{\text{ \ }\alpha
}^{\lambda}\square\Psi_{\lambda\mu}\right\rangle +\frac{1}{4}\left\langle
\square\Psi_{\mu\alpha}\square\Phi\right\rangle . \label{2 termo I}%
\end{equation}

\item Third term:%
\begin{equation}
\eta^{\nu\kappa}\left\langle \Gamma_{\text{ \ }\kappa\nu}^{\rho\left(
1\right)  }\partial^{\beta}C_{\mu\rho\alpha\beta}^{\left(  1\right)
}\right\rangle =-\frac{1}{4}\left\langle \square\Phi\square\Psi_{\mu\alpha
}\right\rangle . \label{3 termo I}%
\end{equation}

\item Fourth term:%
\begin{equation}
\eta^{\nu\kappa}\left\langle \Gamma_{\text{ \ }\kappa\alpha}^{\rho\left(
1\right)  }\partial^{\beta}C_{\mu\nu\rho\beta}^{\left(  1\right)
}\right\rangle =\frac{1}{8}\left[  \left\langle \square\Psi_{\lambda\alpha
}\square\Psi_{\mu}^{\text{ \ }\lambda}\right\rangle +\left\langle \square
\Phi\square\Psi_{\mu\alpha}\right\rangle -\left\langle \partial_{\mu}%
\partial_{\alpha}\Psi_{\lambda\kappa}\square\Psi^{\kappa\lambda}\right\rangle
\right]  . \label{4 termo I}%
\end{equation}

\item Fifth term:%
\begin{equation}
\frac{1}{2}\eta^{\nu\rho}\eta^{\beta\lambda}\left\langle R_{\rho\lambda
}^{\left(  1\right)  }C_{\mu\nu\alpha\beta}^{\left(  1\right)  }\right\rangle
=\frac{1}{8}\left(  \left\langle \square\Psi_{\mu\beta}\square\Psi_{\alpha
}^{\text{ \ }\beta}\right\rangle +\left\langle \partial_{\alpha}\partial_{\mu
}\Psi_{\nu\beta}\square\Psi^{\nu\beta}\right\rangle -\frac{1}{2}\eta
_{\mu\alpha}\left\langle \square\Psi_{\nu\beta}\square\Psi^{\nu\beta
}\right\rangle +\left\langle \square\Psi_{\mu\alpha}\square\Phi\right\rangle
\right)  . \label{5 termo I}%
\end{equation}

\end{itemize}

Substituting these five terms into $\left\langle I_{\mu\alpha}^{\left(
2\right)  }\right\rangle $, we get%
\begin{align}
\left\langle I_{\mu\alpha}^{\left(  2\right)  }\right\rangle  &  =\frac{1}%
{4}\left\langle \square\Phi\square\Psi_{\mu\alpha}\right\rangle -\frac{1}%
{4}\left\langle \square\Psi_{\text{ \ }\alpha}^{\lambda}\square\Psi
_{\lambda\mu}\right\rangle +\frac{1}{4}\left\langle \partial_{\alpha}%
\partial_{\mu}\Psi_{\nu\beta}\square\Psi^{\nu\beta}\right\rangle -\frac{1}%
{16}\eta_{\mu\alpha}\left\langle \square\Psi_{\nu\beta}\square\Psi^{\nu\beta
}\right\rangle \nonumber\\
  &  =-\frac
{1}{4\alpha}\left\langle \partial_{\mu}\Psi_{\nu\beta}\partial_{\alpha}%
\Psi^{\nu\beta}\right\rangle -\frac{1}{4\alpha^{2}}\left\langle \Psi_{\text{
\ }\alpha}^{\lambda}\Psi_{\lambda\mu}\right\rangle -\frac{1}{16\alpha^{2}}%
\eta_{\mu\alpha}\left\langle \Psi_{\nu\beta}\Psi^{\nu\beta}\right\rangle .
\label{I_munu media 1}%
\end{align}

Finally, to obtain the gravitational energy-momentum tensor $t_{\mu\alpha}$,
we substitute Eqs. (\ref{G_munu media 1}), (\ref{H_munu media 1}) and
(\ref{I_munu media 1}) in the expression (\ref{tmualpha}). Thus,%
\[
t_{\mu\alpha}=\frac{c^{4}}{8\pi G}\left[  \frac{1}{4}\left\langle
\partial_{\mu}\tilde{h}_{\nu\beta}\partial_{\alpha}\tilde{h}^{\nu\beta
}\right\rangle -\frac{1}{4}\left\langle \partial_{\mu}\Psi_{\nu\beta}%
\partial_{\alpha}\Psi^{\nu\beta}\right\rangle +\frac{1}{2}\left\langle
\partial_{\mu}\tilde{h}_{\nu\beta}\partial_{\alpha}\Psi^{\nu\beta
}\right\rangle +\frac{3}{2}\left\langle \partial_{\alpha}\Phi\partial_{\mu
}\Phi\right\rangle \right]  .
\]
\\
\\
\\
\\
\\
\\
\\
\\
\\
\\
\\
\\
\end{widetext}

\bibliographystyle{unsrt}
\bibliography{main}

\end{document}